\documentclass[notitlepage,superscriptaddress,nofootinbib,tightenlines,twocolumn]{revtex4-1}
\usepackage{amsmath,verbatim,latexsym,amssymb,indentfirst,mathrsfs,mathtools,amsthm,bbm,bm,hyperref,url,cancel,subcaption}
\usepackage[font=small,labelfont=bf,format=plain,justification=raggedright,singlelinecheck=false]{caption}
\usepackage{verbatim,indentfirst}
\usepackage[title,titletoc]{appendix}

\usepackage{graphicx}
\usepackage{epstopdf}
\newcommand{\caphead}[1]{{\bf #1}}

\renewcommand{\thesection}{\Roman{section}}
\renewcommand{\thesubsection}{\Roman{section} \Alph{subsection}}
\renewcommand{\thesubsubsection}{\Roman{section} \Alph{subsection} \arabic{subsubsection}}
\makeatletter
\def\p@subsection{}
\makeatother
\makeatletter
\def\p@subsubsection{}
\makeatother

\newtheorem{theorem}{Theorem}

\makeatletter
\newcommand\footnoteref[1]{\protected@xdef\@thefnmark{\ref{#1}}\@footnotemark}
\makeatother

\usepackage{soul}

\usepackage{color}
\usepackage[dvipsnames]{xcolor}
\usepackage[normalem]{ulem}

\newcommand{\weak}{{\rm wk}}  
\newcommand{\Fwd}{{\rm F}}  
\newcommand{\Rev}{{\rm R}}  
\newcommand{\I}{{\mathcal{I}}}  
  
\newcommand{\Imag}{{\rm Im}} 
\newcommand{\Real}{{\rm Re}}  

\newcommand*{\KrausV}[2]{ K^{ V,  #1 }_{#2} }  
  
\newcommand*{\KrausVH}[2]{ \hat{K}^{ \hat{V} ,  #1 }_{#2} }   

\newcommand*{\KrausGen}[3]{ K^{ #1 ,  #2 }_{#3} }  

\newcommand*{\KrausGenBB}[2]{ K^{ #1  }_{#2} }  
  
\newcommand*{\KrausGenTwo}[2]{ K^{ #1 }_{#2} }  
\newcommand*{\KrausGenTTwo}[2]{ \tilde{K}^{ #1 }_{#2} }  
\newcommand*{\pV}[1]{ p^V_{ #1} }  
\newcommand*{\pVH}[1]{ p^{ \hat{V} }_{ #1} }  
\newcommand*{\pGen}[2]{ p^{ #1 }_{ #2} }  

\newcommand*{\gV}[1]{ g^V_{ #1 } }  
\newcommand*{\gVH}[1]{ g^{ \hat{V} }_{ #1 } }  

\newcommand*{\gGen}[2]{ g^{ #1 }_{ #2 } }  

\newcommand*{\KrausF}[3]{  \sqrt{ M^{ \Fwd, #1 }_{ #2 , #3 }  }  }  
\newcommand*{\KrausR}[3]{  \sqrt{ M^{ \Rev, #3 }_{ #2, #1 }  }  }  
\newcommand*{\KrausI}[2]{  \sqrt{ M^{ {\rm I} }_{ #1 , #2 }  }  }  
\newcommand*{\KrausII}[1]{  \sqrt{ M^{ {\rm II} }_{ #1 }  }  }  
\newcommand*{\POVMF}[3]{  M^{ \Fwd, #1 }_{ #2 , #3 }  }  
\newcommand*{\POVMR}[3]{  M^{ \Rev, #3 }_{ #2 , #1 }  }  
\newcommand{\ParenK}{{ ( \mathscr{K} ) }}  
\newcommand{\BarK}{{\bar{ \mathscr{K} }}}
\newcommand{\OTOCK}{F^{ ( \BarK ) }}

\newcommand{\W}{W}
\newcommand*{\SumKD}[1]{\tilde{ \mathscr{A} }_{#1}}  
\newcommand*{\ProjW}[1]{\Pi^{ \W }_{#1}}  
\newcommand*{\ProjWt}[1]{\Pi^{ \W(t) }_{#1}}  
\newcommand*{\ProjV}[1]{\Pi^{ V }_{#1}}  
\newcommand*{\ProjVH}[1]{ \hat{\Pi}^{ \hat{V} }_{#1}}  
  
\newcommand*{\ProjWtH}[1]{ \hat{\Pi}^{ \hat{\W}(t) }_{#1}}  
\newcommand*{\DegenW}[1]{ \alpha_{ #1} }  

\newcommand{\Min}{ {\rm min} }   
\newcommand{\Max}{ {\rm max} }   
\newcommand{\inter}{ {\rm int} }   

\newcommand{\Sh}{{\rm Sh}}   
\newcommand{\vN}{{\rm vN}}   
\def\const{ {\rm const.} }   
\newcommand{\Tr}{{\rm Tr}}   
\def\id{\mathbbm{1}}   
\newcommand{\kB}{k_\mathrm{B}}  

\newcommand{\Hil}{\mathcal{H}}  
\newcommand{\sys}{{\rm s}}  
\newcommand{\detect}{{\rm d}}  
\newcommand{\Sites}{N}  
\newcommand{\Dim}{d}   

\newcommand{\LParen}{ \bm{(} }
\newcommand{\RParen}{ \bm{)} }

\newcommand*{\Set}[1]{\left\{  #1  \right\}}
\newcommand*{\Verts}[1]{\left\lvert  #1  \right\rvert}

\renewcommand\th{ {\rm th} }

\newcommand*{\bra}[1]{\langle #1\rvert}
\newcommand*{\ket}[1]{\lvert #1 \rangle}
\newcommand*{\braket}[2]{\langle #1 \lvert #2 \rangle}
\newcommand*{\ketbra}[2]{\lvert #1 \rangle\!\langle #2 \rvert}
\newcommand*{\expval}[1]{\left\langle  #1  \right\rangle}

\begin{document}
\title{Entropic uncertainty relations for quantum information scrambling}
\author{Nicole~Yunger~Halpern}
\email{Current address and email: 
Harvard-Smithsonian ITAMP, 
60 Garden St., MS 14,
Cambridge, MA 02138, USA,
nicoleyh@g.harvard.edu}
\affiliation{Institute for Quantum Information and Matter, California Institute of Technology, Pasadena, CA 91125, USA}
\affiliation{Kavli Institute for Theoretical Physics, University of California, Santa Barbara, CA 93106, USA}
\author{Anthony~Bartolotta}
\email{abartolo@theory.caltech.edu}
\affiliation{Walter Burke Institute for Theoretical Physics, California Institute of Technology, Pasadena, CA 91125, USA}
\author{Jason~Pollack}
\email{jpollack@phas.ubc.ca}
\affiliation{Department of Physics and Astronomy, University of British Columbia, Vancouver, BC V6T 1Z1, Canada}
\date{\today}

%
%
\begin{abstract}
How violently do two quantum operators disagree? Different fields of physics feature different measures of incompatibility: (i) In quantum information theory, entropic uncertainty relations constrain measurement outcomes. (ii) In condensed matter and high-energy physics, the out-of-time-ordered correlator (OTOC) signals scrambling, the spread of information through many-body entanglement. We unite these measures, proving entropic uncertainty relations for scrambling. The entropies are of distributions over weak and strong measurements' possible outcomes. The weak measurements ensure that the OTOC quasiprobability (a nonclassical generalization of a probability, which coarse-grains to the OTOC) governs terms in the uncertainty bound. The quasiprobability causes scrambling to strengthen the bound in numerical simulations of a spin chain. This strengthening shows that entropic uncertainty relations can reflect the type of operator disagreement behind scrambling. Generalizing beyond scrambling, we prove entropic uncertainty relations satisfied by commonly performed weak-measurement experiments. We unveil a physical significance of weak values (conditioned expectation values): as governing terms in entropic uncertainty bounds.
\end{abstract}

{\let\newpage\relax\maketitle}

%
%

How incompatible are two quantum operators,
 $\hat{V}$ and $\hat{\W}(t)$?
Two species of quantum physicist answer with two different measures.
Today's pure quantum information (QI) theorist
checks uncertainty relations cast in terms of entropies~\cite{Everett_57_Relative,Hirschman_57_Note,Beckner_75_Inequalities,Bialynicki_Birula_75_Uncertainty,Deutsch_83_Uncertainty,Kraus_87_Complementary,Maassen_88_Generalized,Ghirardi_03_Optimal,Christandl_05_Uncertainty,DeVicente_08_Improved,Renes_09_Conjectured,Berta_10_Uncertainty,Coles_12_Uncertainty,Tomamichel_12_Framework,Coles_17_Entropic,Wehner_10_Entropic}.
The greater the uncertainty bounds, 
the worse the operators' disagreement.

The second species---the 
condensed-matter or high-energy physicist---studies 
the following setup:
Consider a strongly coupled quantum many-body system.
Examples include an interacting spin chain and
the boundary dual of a gravitational theory.
The Hamiltonian, $\hat{H}$, couples the subsystems
and generates the time-evolution operator
$\hat{U}  :=  e^{ - i \hat{H} t }$.
Let $\hat{V}$ and $\hat{\W}$ denote Hermitian and/or unitary operators
localized on far-apart subsystems.
Examples include Pauli operators acting
on opposite sides of the spin chain.
In the Heisenberg picture,
the interactions delocalize $\hat{\W}$ to
$\hat{\W}(t)  :=  \hat{U}^\dag  \hat{\W}  \hat{U}$.
The support of $\hat{\W}(t)$ comes to overlap the support of $\hat{V}$;
the operators cease to agree.
The \emph{out-of-time-ordered correlator} (OTOC)
quantifies this disagreement,
as well as quantum chaos and scrambling~\cite{LarkinO_69,Lashkari_13_Towards,Kitaev_15_Simple,Shenker_Stanford_14_BHs_and_butterfly,Shenker_Stanford_14_Multiple_shocks,Roberts_15_Localized_shocks,Roberts_Stanford_15_Diagnosing,Maldacena_15_Bound,Haehl_16_Schwinger_I,Hosur_16_Chaos,Bohrdt_16_Scrambling,Maldacena_16_Remarks,Aleiner_16_Microscopic,Roberts_16_Lieb,Huang_17_MBL_OTOC,Swingle_17_MBL_OTOC,Fan_17_MBL_OTOC,He_17_MBL_OTOC}.
QI \emph{scrambles} upon spreading across a system
via many-body entanglement.

Entropic uncertainty relations and OTOCs
occupy disparate subfields,
but both quantify operator disagreement.
We unite these quantifications,
proving entropic uncertainty relations for QI scrambling
(Theorem~\ref{theorem:OTOC_thm_Toma}).
These relations make precise the extent to which 
scrambling drives operators away from compatibility.
We then evaluate the uncertainty relations 
in numerical simulations of nonintegrable spin chains.
The uncertainty bounds tighten when the system scrambles.
Hence entropic uncertainty relations can reflect
the type of operator incompatibility behind scrambling.
The bounds tighten because they contain 
the quasiprobability behind the OTOC~\cite{NYH_17_Jarzynski,NYH_18_Quasi,Alonso_18_Out}.
\emph{Quasiprobabilities} resemble probabilities
but can behave nonclassically,
assuming negative and nonreal values.
The OTOC has been shown to equal an average
over a quasiprobability distribution.
Recent studies have uncovered
several theoretical and experimental applications
of the OTOC quasiprobability~\cite{NYH_17_Jarzynski,NYH_18_Quasi,Alonso_18_Out}.
We present a new application:
The quasiprobability enables entropic uncertainty relations
to reflect the same operator disagreement as the OTOC.

Our entropic uncertainty relations generalize in two ways.
First, they extend from the famous four-point OTOC
to arbitrarily high-order, arbitrarily out-of-time-ordered correlators.
Such OTOCs reflect later, subtler stages 
of scrambling and equilibration~\cite{Roberts_16_Chaos,Haehl_17_Classification,Haehl_17_Thermal,NYH_18_Quasi,Dressel_18_Strengthening,Haehl_18_Fine}.
Second, our entropic uncertainty relations generalize
beyond many-body systems that scramble.
Quasiprobabilities govern weak measurements, 
which barely disturb the measured system~\cite{Tamir_13_Introduction},
similarly to how probabilities govern strong measurements.
Averaging a common quasiprobability yields 
a \emph{weak value}, a conditioned expectation value.
Weak values' physical significances have been debated~\cite{Aharonov_88_How,Dressel_14_Understanding}.
We present a new physical significance: 
Weak values, as well as quasiprobabilities, 
govern first-order-in-the-weak-coupling terms
in entropic uncertainty bounds (Theorem~\ref{theorem:OTOC_free}).
These bounds govern weak-measurement experiments 
undertaken routinely.

The rest of this paper is organized as follows.
We briefly overview entropic uncertainty relations and OTOCs
in Section~\ref{sec_Background}.
In Sec.~\ref{sec_Intuit}, we reason intuitively to
the form that entropic uncertainty relations for scrambling
should assume.
This intuition is made rigorous in Sec.~\ref{sec_Formalization}.
Numerical simulations of a spin chain
illustrate Theorem~\ref{theorem:OTOC_thm_Toma} 
in Sec.~\ref{sec:Numerics}.
We generalize to higher-point OTOCs in Sec.~\ref{sec:K_OTOCs},
then to weak values beyond scrambling in Sec.~\ref{sec:Beyond}.

\section{Background}
\label{sec_Background}

Here, we review the two measures of operator disagreement,
entropic uncertainty relations and OTOCs.

\subsection{Entropic uncertainty relations}
\label{sec_EUR_Review}

Heisenberg captured the complementarity
of position and momentum in~\cite{Heisenberg_27_Uber}.
Kennard concretized this complementarity in
the first uncertainty relation~\cite{Kennard_27_Zur}.
Robertson proved the uncertainty relation 
featured in many textbooks~\cite{Robertson_29_Uncertainty}:
\begin{align}
   \label{eq:Robertson}
   \Delta \hat{A} \, \Delta \hat{B}  
   \geq  \frac{1}{ 2}  \, 
   | \langle [ \hat{A},  \hat{B} ] \rangle |  \, .
\end{align}
We have set $\hbar$ to one.
$\hat{A}$ and $\hat{B}$ denote observables
defined on a Hilbert space $\Hil$.
The expectation value $\expval{ . }$ is evaluated on 
a state $\ket{ \psi }  \in  \Hil$.
The standard deviation $\Delta \hat{A}  
:=  \sqrt{ \langle \hat{A}^2 \rangle  -  \langle \hat{A} \rangle^2 }$
quantifies the spread in the possible outcomes of
a measurement of $\hat{A}$.

The standard deviations have provoked objections
(e.g.,~\cite{Deutsch_83_Uncertainty}).
For example, consider relabeling the eigenvalues 
$a$ of $\hat{A}$.
Relabeling should not change the operators' compatibility,
but $\Delta \hat{A}$ can skyrocket.
Stripping the $a$'s off of $\Delta \hat{A}$ leaves
a function of probabilities:
Denote by $p_{a}$ the probability that 
a measurement of $\hat{A}$ yields $a$.
On probability distributions $\Set{ p_{a} }$ are defined entropies.
Entropies, the workhorses of information theory, 
quantify the optimal rates at which information-theoretic
and thermodynamic tasks can be performed~\cite{Shannon_48_Mathematical,NielsenC10,Tomamichel_16_Quantum,Wilde_17_Quantum}.
Entropies replace standard deviations in modern uncertainty relations~\cite{Everett_57_Relative,Hirschman_57_Note,Beckner_75_Inequalities,Bialynicki_Birula_75_Uncertainty,Deutsch_83_Uncertainty,Kraus_87_Complementary,Maassen_88_Generalized,Ghirardi_03_Optimal,Christandl_05_Uncertainty,DeVicente_08_Improved,Renes_09_Conjectured,Berta_10_Uncertainty,Coles_12_Uncertainty,Tomamichel_12_Framework,Coles_17_Entropic,Wehner_10_Entropic}.

The \emph{Maassen-Uffink relation} exemplifies
entropic uncertainty relations~\cite{Maassen_88_Generalized}:
\begin{align}
   \label{eq:MU}
   H ( \hat{A} )  +  H ( \hat{B} )  
   \geq  - \log c  \, .  
\end{align}
The Shannon entropy is defined as
$H( \hat{A} )  :=  -  \sum_{a}  p_{a}  \log  p_{a}$.
The \emph{maximum overlap} $c$ is defined in terms of
the eigendecompositions
\begin{align}
   \label{eq:Decomp_A}
   & \hat{A}  =  \sum_a  a   \ketbra{a}{a}  
   \quad \text{and} \quad  
   \hat{B}  =  \sum_b  b  \ketbra{b}{b}  
\end{align}
as
\begin{align}
   \label{eq:Overlap_Def}
   c  :=  \max_{a, b} 
   | \braket{a}{b}  |^2  \, .
\end{align}
Hence the bound~\eqref{eq:MU} is independent of 
the eigenvalues $a$, as desired.
The bound is tight if $c$ is small.
$c$ is smallest when the eigenbases are 
\emph{mutually unbiased}:
$| \braket{a}{b}  |
=  \frac{1}{  \sqrt{  \Dim  }  }$,
wherein $\Dim  :=  \dim ( \Hil )$ denotes 
the Hilbert space's dimensionality.
For example, the Pauli operators 
$\hat{\sigma}^x$ and $\hat{\sigma}^z$  
have mutually unbiased eigenbases.
If you prepare any eigenstate of $\hat{\sigma}^x$,
then measure $\hat{\sigma}^z$,
you have no idea which outcome will obtain.
Hence $\hat{\sigma}^x$ and $\hat{\sigma}^z$ are said
to fail \emph{maximally} to commute.
Entropic uncertainty relations have applications
to many topics in quantum theory, including
quantum correlations, steering, coherence, and wave-particle duality
(see~\cite{Coles_17_Entropic} and references therein).

\subsection{Out-of-time-ordered correlators}
\label{sec_OTOC_Review}

OTOCs reflect chaos and QI spreading in quantum many-body systems.
Settings range from ultracold atoms and trapped ions
to holographic black holes (e.g.,~\cite{LarkinO_69,Lashkari_13_Towards,Kitaev_15_Simple,Shenker_Stanford_14_BHs_and_butterfly,Shenker_Stanford_14_Multiple_shocks,Roberts_15_Localized_shocks,Roberts_Stanford_15_Diagnosing,Maldacena_15_Bound,Haehl_16_Schwinger_I,Hosur_16_Chaos,Bohrdt_16_Scrambling,Maldacena_16_Remarks,Garttner_16_Measuring,Aleiner_16_Microscopic,Roberts_16_Lieb,Huang_17_MBL_OTOC,Swingle_17_MBL_OTOC,Fan_17_MBL_OTOC,He_17_MBL_OTOC}).
Let $\Hil$ denote a quantum many-body system's Hilbert space.
Let $\hat{\rho} \in \mathcal{D} ( \Hil )$ denote
an arbitrary state of the system.
$\mathcal{D} ( \Hil )$ denotes the space of density operators,
or trace-one positive-semidefinite linear operators,
defined on $\Hil$.
The OTOC has the form
\begin{align}
   \label{eq:OTOC_Def}
   F(t)  :=  \expval{ \hat{\W}^\dag (t)  \hat{V}^\dag  \hat{\W}(t)  \hat{V} }
   \equiv  \Tr \left(  \hat{\W}^\dag (t)  \hat{V}^\dag  \hat{\W}(t)  \hat{V}  
   \hat{\rho}  \right)  ,
\end{align}
for unitary and/or Hermitian $\hat{V}$ and $\hat{\W}$ 
localized far apart.
The OTOC forms the nontrivial component of
\begin{align}
   \label{eq:Com_Squared}
   \expval{ | [  \hat{\W} (t),  \hat{V} ] |^2 }  \, .
\end{align}
This magnitude-squared commutator equals $2 [ 1 - F(t) ]$
if $\hat{V}$ and $\hat{W}$ are unitary (e.g., Pauli operators).

Several pieces of evidence imply that the OTOC signals chaos.
We review a semiclassical argument about the butterfly effect:
Classical chaos hinges on sensitivity to initial perturbations.
Consider initializing a classical double pendulum
at a phase-space point $\mathbf{P}$
with a strong kick.
Let the pendulum begin another trial
at a nearby point 
$\mathbf{P} + \bm{\varepsilon}$.
The pendulum follows different phase-space trajectories
in the two trials.
The trajectories diverge exponentially,
as quantified with a Lyapunov exponent.

The OTOC captures a similar divergence.
Let us construct two protocols that differ 
largely by an initial perturbation.
The system could consist of
an $\Sites$-site chain of spin-$\frac{1}{2}$ 
degrees of freedom, or \emph{qubits}.
Suppose that $\hat{\rho} = \ketbra{\psi}{\psi}$ is pure.
Protocol I consists of (i) preparing the system
in $\ket{ \psi }$,
(ii) perturbing the system with a local $\hat{V}$
(as by flipping spin 1 with $\hat{\sigma}^x_1$),
(iii) evolving the system under a nonintegrable Hamiltonian,
(iv) perturbing with a local $\hat{\W}$ 
(such as the final spin's $\hat{\sigma}^z_\Sites$),
and (v) evolving the system backward, under $\hat{U}^\dag$.
This protocol prepares
$\ket{ \psi_{\rm I} }  :=  \hat{\W}(t)  \hat{V}  \ket{ \psi }$.

Following protocol II, one prepares $\ket{ \psi }$
and skips the initial $\hat{V}$.
The system evolves forward under $\hat{U}$, 
is perturbed with $\hat{\W}$,
and reverse-evolves under $\hat{U}^\dag$.
Only afterward does $\hat{V}$ perturb the system.
Protocol II prepares $\ket{ \psi_{\rm II} }  :=  \hat{V}  \hat{\W}(t) \ket{ \psi }$.

How much does the initial $\hat{V}$ perturbation 
affect the system's final state?
The answer manifests in the overlap
\begin{align}
   \label{eq:Overlap_OTOC}
   | \braket{ \psi_{\rm II} }{ \psi_{\rm I} } |
   = | F(t) |
   \sim  1  -  \frac{  e^{ \lambda t }  }{ \Sites }  \, .
\end{align}
Nonlocal systems, such as the Sachdev-Ye-Kitaev (SYK) model~\cite{Sachdev_93_Gapless,Kitaev_15_Simple,Polchinski_16_Spectrum,Maldacena_16_Remarks},
obey the final relation.
[In local systems, $F(t)$ decays polynomially.]
The relation holds during a time window around
the \emph{scrambling time}, $t_*$.
The  Lyapunov-type exponent $\lambda$
controls the exponential decay.
Hence $F(t)$ reflects a Lyapunov-type divergence
reminiscent of classical-chaotic sensitivity to initial perturbations.

Smallness of $F(t)$ tends to reflect 
highly nonlocal entanglement.
After $t_*$, no local probe $\hat{V}$ can recover
information about any earlier, initially local perturbation $\hat{\W}$.
This many-body nonlocality is called \emph{scrambling}~\cite{Brown_13_Scrambling,Hosur_16_Chaos}.

\section{Intuitive construction of entropic uncertainty relations 
for quantum information scrambling}
\label{sec_Intuit}

Uncertainty relations and OTOCs,
reflecting quantum operator disagreement
in different subfields, cry out for unification.
But how can one form an uncertainty relation for scrambling?
One might try substituting $\hat{A} = \hat{V}$ and $\hat{B} = \hat{\W}(t)$  
into the uncertainty relation~\eqref{eq:MU}.
But the bound would bear no signature of scrambling.
Moreover, simulations imply,
simple choices of $\hat{V}$ and $\hat{\W}(t)$ eigenbases
fail to become mutually unbiased after $t_*$~\cite{NYH_18_Quasi}.

A clue suggests how entropic uncertainty relations for scrambling
may be realized:
The entropic inequality~\eqref{eq:MU} replaced 
the textbook inequality~\eqref{eq:Robertson}.
Inequality~\eqref{eq:Robertson} contains one commutator.
The OTOC appears in a commutator's squared magnitude
[Eqs.~\eqref{eq:OTOC_Def} and~\eqref{eq:Com_Squared}].
Hence ``squaring'' Ineq.~\eqref{eq:MU}, in some sense,
might yield an entropic uncertainty relation for scrambling.

How might this ``squaring'' manifest?
In the left-hand side (LHS) of Ineq.~\eqref{eq:MU},
each entropy $H$ depends on one operator, $\hat{A}$ or $\hat{B}$.
Imagine ``doubling'' each operator
by replacing it with two operators.
The two operators suited to scrambling
are $\hat{V}$ and $\hat{\W}(t)$.
We therefore envision an entropy 
$H \LParen \hat{V} \hat{\W}(t) \RParen$
defined in terms of a measurement of $\hat{V}$
followed by a measurement of $\hat{\W}(t)$.
This replacement for $H( \hat{A} )$ must differ from
the replacement for $H ( \hat{B} )$,
but the OTOC contains only two local operators.
We therefore reverse the measurements:
$H( \hat{B} )  \to
H \LParen \hat{\W}(t) \hat{V} \RParen$.
The reversal mirrors the OTOC's semiclassical interpretation, Eq.~\eqref{eq:Overlap_OTOC}.


How can the right-hand side (RHS) of Ineq.~\eqref{eq:MU} 
be ``squared''?
$c$ equals a product of two inner products.
``Squaring'' $c$ creates a product of four inner products,
or the trace of four outer products
$\ketbra{ \ldots }{ \ldots }$.
Outer products generalize to projectors $\Pi$.
Hence a trace of a product of four projectors,
$\Tr ( \Pi  \,  \Pi  \,  \Pi  \,  \Pi )$,
should appear in an entropic uncertainty bound for scrambling.
Such a trace is known to characterize scrambling.
It forms the \emph{quasiprobability behind the OTOC}~\cite{NYH_17_Jarzynski,NYH_18_Quasi,Alonso_18_Out}.

Quasiprobability distributions represent quantum states
as probability distributions
represent classical statistical-mechanical states.
Like probabilities, quasiprobabilities are normalized to one.
Yet quasiprobabilities violate axioms of probability theory,
such as nonnegativity and reality.
Such nonclassical behaviors can signal nonclassical physics, 
such as the capacity for superclassical computation~\cite{Gross_05_Computational,Spekkens_08_Negativity,Veitch_12_Negative,Howard_14_Contextuality}.

The OTOC equals an average over a quasiprobability distribution
defined as follows~\cite{NYH_17_Jarzynski,NYH_18_Quasi}.
The OTOC operators eigendecompose as
\begin{align}
   \label{eq_VW_Decomp}
   \hat{V}  =  \sum_{v_\ell}  v_\ell \,
   \ProjVH{v_\ell}
   \quad \text{and} \quad
   \hat{\W}(t)  =  \sum_{w_m}  w_m \,
   \ProjWtH{w_m}  \, .
\end{align}
In the spin-chain example, the eigenvalues
$v_\ell, w_m = \pm 1$.
The projector $\ProjVH{v_\ell}$ projects onto
the eigenvalue-$v_\ell$ eigenspace of $\hat{V}$.
$\ProjWtH{w_m}$ is defined analogously.
Consider substituting from Eqs.~\eqref{eq_VW_Decomp}
into the OTOC definition~\eqref{eq:OTOC_Def}.
Factoring out the sums and the eigenvalues yields\footnote{
The index list $(v_1, w_1, v_2, w_2)$ here is equivalent to
the index list $(v_1, w_2, v_2, w_3)$ in~\cite{NYH_17_Jarzynski,NYH_18_Quasi}.}
\begin{align}
   \label{eq:Decomp_F_Coarse}
   F(t)  =  \sum_{ v_1, w_1, v_2, w_2 }
   v_1 w_1 v_2^* w_2^*  \:
   \SumKD{ \hat{\rho} } ( v_1, w_1, v_2, w_2 )  \, .
\end{align}
The OTOC equals an average over the OTOC quasiprobability,
\begin{align}
   \label{eq:SumKD_Def}
   \SumKD{ \hat{\rho} } ( v_1, w_1, v_2, w_2 )  
   & :=  \Tr \left(  \ProjWtH{w_2}  \ProjVH{v_2}  
   \ProjWtH{w_1}  \ProjVH{v_1}  \hat{\rho}  \right)  \, .
\end{align}

The quasiprobability forms a distribution $\{ \SumKD{ \hat{\rho} } \}$.
This set of numbers contains more information than the OTOC,
which follows from coarse-graining.
$\SumKD{ \hat{\rho} }$ concretizes the relationship 
between scrambling and nonequilibrium statistical mechanics~\cite{NYH_17_Jarzynski},
informs schemes for measuring the OTOC experimentally~\cite{NYH_17_Jarzynski,NYH_18_Quasi,Dressel_18_Strengthening},
distinguishes scrambling from decoherence in
measurements of open-system OTOCs~\cite{Alonso_18_Out},
and underlies a quantum advantage in metrology~\cite{Shukur_19_Contextuality}.
This paper presents a new application of the OTOC quasiprobability:
$\SumKD{ \hat{\id} }$ governs terms in 
the entropic uncertainty bound for scrambling.
The quasiprobability tightens the bound when the system scrambles.
We evaluate the quasiprobability on the identity operator $\hat{\id}$
because entropic uncertainty bounds 
cannot depend on any state $\hat{\rho}$.
Uncertainty relations require, moreover, 
that eigenvalues be ``stripped off'' of operators.
$\SumKD{ \hat{\id} }$ follows from 
stripping the eigenvalues off the OTOC,
by Eq.~\eqref{eq:Decomp_F_Coarse}.

We can predict the form of the uncertainty-bound term 
that will contain $\SumKD{ \hat{\id} }$.
Quasiprobabilities can be measured via weak measurement:
An interaction Hamiltonian couples a detector to the system.
A small coupling constant $g$ governs the interaction.
The measurement disturbs the measured state
at high order in $g$.
From weak and strong measurements of $\hat{V}$ and $\hat{\W}(t)$,
the OTOC quasiprobability can be inferred experimentally~\cite{NYH_17_Jarzynski,NYH_18_Quasi,Dressel_18_Strengthening}.
$\SumKD{ \hat{\rho} }$ is extracted from the data
through a high-order term.
$\SumKD{ \hat{\id} }$ should therefore appear 
in a high-order-in-$g$ term in our entropic uncertainty bound.

The uncertainty relation's RHS contains $g$
only if the LHS involves weak measurements.
Consider measuring $\hat{V}$ weakly, then $\hat{\W}(t)$ strongly.
Each possible pair $(v_\ell, w_m)$ of outcomes
has some probability of obtaining.
On this probability, we propose to define the entropy
$H \LParen \hat{V} \hat{\W}(t) \RParen$.
$H \LParen \hat{\W}(t) \hat{V} \RParen$
should be defined similarly.

Let us summarize our intuitive reasoning.
Entropic uncertainty relations for scrambling
should have the form
\begin{align}
   \label{eq_Unc_Form} & 
   H \LParen \hat{V} \hat{\W}(t) \RParen
   +  H \LParen \hat{\W}(t) \hat{V} \RParen
   \geq  g^{k-1} ( \text{classical factor} )
   \nonumber \\ & \qquad
   + g^k (\const) \SumKD{ \hat{\id} } ( v_1, w_1, v_2, w_2 )
   + O ( g^{k+1} ) .
\end{align}
The exponent $k \geq 2$.
$H \LParen \hat{V} \hat{\W}(t) \RParen$
quantifies the uncertainty about the outcomes
that follow from preparing an arbitrary $\hat{\rho}$,
measuring $\hat{V}$ weakly, 
and then measuring $\hat{\W}(t)$ strongly.
$H \LParen \hat{\W}(t) \hat{V} \RParen$
results from reversing the measurement protocol.
Having constructed expectations via intuition,
we now prove them.

\section{Formalization of entropic uncertainty relations
for quantum information scrambling}
\label{sec_Formalization}

We introduce the setup in Sec.~\ref{sec_Setup}
and formalize our measurements
in Sec.~\ref{sec:POVM}.
The measurements yields outcomes distributed
according to probability distributions
on which are defined entropies in Sec.~\ref{sec:Entropies}.
Our main result (Theorem~\ref{theorem:OTOC_thm_Toma})
is presented in Sec.~\ref{sec:Main_thm_and_analysis}
and analyzed in Sec.~\ref{sec:Analysis_MainThm}.

\subsection{Setup}
\label{sec_Setup}

We continue to focus on a quantum many-body system
illustrated with a chain of $\Sites$ qubits.
To simplify notation, we omit hats from operators.
Many-body quantities are defined as in the introduction:
the Hilbert space $\Hil$,
its dimensionality $\Dim$,
the arbitrary state $\rho \in \mathcal{D} ( \Hil )$,
the Hamiltonian $H$,
the time-evolution unitary $U$,
the local operators $V$ and $\W$
(illustrated with $\sigma^z_1$ and $\sigma^z_\Sites$),
the Heisenberg-picture $\W(t)$,
the projectors $\ProjV{v_\ell}$ and $\ProjWt{w_m}$,
the eigenvalues $v_\ell$ and $w_m$,
the OTOC $F(t)$,
and the OTOC quasiprobability $\SumKD{\rho}$.

The Hilbert space $\Hil$ is assumed to be discrete,
in accordance with~\cite{Tomamichel_12_Framework,Krishna_01_Entropic},
whose results we use.
Continuous-variable systems are addressed 
in Sec.~\ref{sec:Discussion}.
We emphasize nonintegrable, nonlocal Hamiltonians.
We assume that $V$ and $\W$ are Hermitian, for simplicity,
but the results generalize: Each of $V$ and $\W$ 
can be Hermitian and/or unitary~\cite{NYH_17_Jarzynski,NYH_18_Quasi}.
If $V$ is unitary but not Hermitian, for example,
measurements of $V$ are replaced with measurements of
the Hermitian generator of $V$.

\subsection{Formalization of measurements}
\label{sec:POVM}

A sequence of $V$ and $\W(t)$ measurements
forms a generalized measurement.
Generalized measurements are formalized, in QI theory,
with \emph{positive operator-valued measures} (POVMs)~\cite{NielsenC10}.
A POVM $\Set{  M_x }$ consists of positive operators
$M_x  >  0$ that obey the completeness condition
$\sum_x  M_x^\dag M_x  =  \id$.
$x$ labels the outcomes.

POVMs replace measurements of observables $A$ and $B$
in generalized entropic uncertainty relations~\cite{Krishna_01_Entropic,Tomamichel_12_Framework}.
We adapt the formalism used by Tomamichel~\cite{Tomamichel_12_Framework},
for concreteness and for ease of comparison with a standard reference.
In~\cite{Tomamichel_12_Framework} appear POVMs
illustrated with measurements of observables.

These general POVMs manifest, in the context of scrambling, as follows.
We label as ``the forward measurement'' 
a weak measurement of $V$,
followed by a projective measurement of $\W(t)$.
We use the term ``weak measurement of $V$''
as in~\cite{NYH_18_Quasi}:
A projector $\ProjV{v_1}$ is effectively measured weakly.
One can effectively measure a qubit system's $\ProjV{v_1}$
by, e.g., coupling the detector to $V$
and calibrating the detector appropriately.
The experimenter chooses the value of $v_1$;
the choice directs the calibration.
See Sec.~\ref{sec:Spin_Chain_Set_up} 
and~\cite[Sec.~I D 4]{NYH_18_Quasi} 
for example implementations.
The reverse process constitutes the second POVM,
for a definition of ``reverse'' that we concretize
after formalizing the weak measurement.

To measure $\ProjV{v_\ell}$ weakly,
one prepares a detector in a state $\ket{D}$.
The system's $\ProjV{v_\ell}$ is coupled weakly
to a detector observable,
via an interaction unitary $V_\inter$.
A detector observable is measured projectively,
yielding an outcome $j_\ell$.

The weak measurement induces dynamics
modeled with \emph{Kraus operators}~\cite{NielsenC10,Preskill_15_Ch3}.
Kraus operators represent the system-of-interest evolution
effected by a coupling to an ancilla,
which effectively measures the system:
\begin{align}
   \label{eq:Kraus}
   \KrausV{ v_\ell }{ j_\ell }
   =  \bra{ j_\ell }  V_\inter  \ket{ D }
   =  \sqrt{  \pV{ j_\ell }  }  \:  \id
   +  \gV{ j_\ell }  \,  \ProjV{ v_\ell }  \, .
\end{align}
The operators satisfy the completeness relation
$\sum_{ j_\ell }  \big(  \KrausV{ v_\ell }{ j_\ell }  \big)^\dag
\KrausV{ v_\ell }{ j_\ell }  =  \id$.
Let $\rho$ temporarily denote
the system's precoupling state.
The detector has a probability 
$\Tr  \big(  \KrausV{ v_\ell }{ j_\ell }  \rho  
\big[  \KrausV{ v_\ell }{ j_\ell }  \big]^\dag  \big)$
of registering the outcome $j_\ell$.
The outcome-dependent $\gV{j_\ell}  \in  \mathbb{C}$ quantifies
the interaction strength.
The experimenter can tune $\gV{j_\ell}$,
whose smallness reflects the measurement's weakness:
$\Verts{  \gV{j_\ell}  }   \ll  1$.
We refer to various constants $\gV{j_\ell}$ as $g$'s.

Imagine strongly measuring the detector observable
without having coupled the detector to the system.
The outcome $j_\ell$ has a probability $\pV{j_\ell}$
of obtaining.
We invoke Kraus operators' unitary equivalence~\cite{Preskill_15_Ch3}
to ensure that $\pV{j_\ell}  \in  \mathbb{R}$.

The forward POVM $\Set{ \POVMF{ v_1 }{ j_1 }{ w_1 } }$
is defined through the composite Kraus operators
\begin{align}
   \label{eq:Fwd_POVM}
   \KrausF{ v_1 }{ j_1 }{ w_1 }  
   :=  \ProjWt{w_1}     \KrausV{v_1}{ j_1 }  \, .
\end{align}
Recall that $\ProjWt{w_1}$ projects onto
the $w_1$ eigenspace of $\W(t)$.
Each POVM element has the form 
$\left(  \KrausF{ v_1 }{ j_1 }{ w_1 }  \right)^\dag
\KrausF{ v_1 }{ j_1 }{ w_1 } $.

The reverse POVM, $\Set{  \POVMR{ w_2 }{ j_2 }{ v_2 }  }$,
is defined through the composite Kraus operators\footnote{
Ending the protocol with a weak measurement
might disconcert measurement theorists.
But this reverse protocol captures 
the OTOC's forward-and-reverse spirit 
[Eq.~\eqref{eq:Overlap_OTOC}],
as explained earlier.}
\begin{align}
   \label{eq:Rev_POVM}
   \KrausR{ w_2 }{ j_2 }{ v_2 }
   & :=   \left(  \ProjWt{w_2}   \KrausV{ v_2 }{ j_2 }  \right)^\dag
   =   \left(  \KrausV{ v_2 }{ j_2 }  \right)^\dag
   \ProjWt{w_2}  \, .
\end{align}
To round out the reversal,
we not only swap the $V$ measurement with the $\W(t)$,
but also Hermitian-conjugate.
Conjugation negates imaginary numbers.
It represents, e.g., the time-reversal of magnetic fields.

Let us clarify which variables are chosen and which vary randomly.
$w_1$ is a random outcome
whose value varies from realization to realization
of the forward POVM.
$w_2$ is a random outcome
whose value varies from realization to realization 
of the reverse POVM.
The experimentalist chooses
the values of $v_1$ and $v_2$.
Though a forward trial's $v_1$ and $w_1$ 
can differ from a reverse trial's $v_2$ and $w_2$,
both protocols' measurements [of $V$ and of $\W(t)$]
are essentially the same.

%
%
%
\subsection{Entropies}
\label{sec:Entropies}

Consider preparing the system in the state $\rho$, 
then measuring the forward POVM,
$\Set{  \POVMF{ v_1 }{ j_1 }{ w_1 }  }$.
One prepares a detector in some fiducial state.
Some detector observable is effectively coupled to
the system's $\ProjV{v_1}$.
Then, some detector observable 
couples to a classical\footnote{
``Classical'' means, here, that 
the register can occupy only quantum states
representable by density matrices
diagonal with respect to a fixed basis.}
register.
The register records an outcome $j_1$. 
Next, the system's $\W(t)$ couples to another classical register.
This register records the outcome $w_1$.

The two-register system ends in the state
\begin{align}
   \rho_{ \Fwd }  & :=
   \sum_{ j_1 , w_1 }
   \Tr  \left(  \KrausF{ v_1 }{ j_1 }{ w_1 }  \rho  
                  \KrausF{ v_1 }{ j_1 }{ w_1 }^\dag   \right)
   \ketbra{ j_1 }{ j_1 }  \otimes
   \ketbra{ w_1 }{ w_1 }  \, .
\end{align}
The eigenvalues,
$\Tr  \Big(  \KrausF{ v_1 }{ j_1 }{ w_1 }  \rho  
                  \KrausF{ v_1 }{ j_1 }{ w_1 }^\dag   \Big)$,
form a probability distribution over 
the possible pairs $( j_1, w_1 )$
of measurement outcomes.
Entropies of the distribution equal entropies of $\rho_\Fwd$.\footnote{
In defining the entropies,
we mostly follow Tomamichel's conventions~\cite{Tomamichel_12_Framework}.
Yet we assume that all states $\sigma$ are normalized:
$\Tr ( \sigma ) = 1$.}

The order-$\alpha$ R\'enyi entropy of 
a quantum state $\sigma$ is 
\begin{align}
   H_\alpha ( \sigma )  :=
   \frac{1}{ 1 - \alpha }  \,  
   \log  \LParen  \Tr  
   \left(  \sigma^\alpha  \right)  \RParen  \, .
\end{align}
We choose for all logarithms to be base-2, 
following~\cite{Tomamichel_12_Framework}.
The von Neumann entropy is
\begin{align}
   \label{eq:H_vN_Def}
   H_\vN ( \sigma )  
   & =  \lim_{ \alpha \to 1 }  H_\alpha ( \sigma )  
   =  - \Tr ( \sigma \log \sigma )  \, .
\end{align}
The \emph{min entropy} is defined as
\begin{align}
   \label{eq:HMin_Def}
   H_\Min ( \sigma )  
   & :=  H_\infty ( \sigma )
   :=  \lim_{ \alpha \to \infty }  H_\alpha ( \sigma )  \\
   & =  \sup \{ \lambda \in \mathbb{R}  \, :  \,
   \sigma  \leq  2^{ - \lambda }  \id  \}  \\
   & =  - \log ( p_\Max )   \, .
\end{align}
$p_\Max$ denotes the greatest eigenvalue of $\sigma$.

The \emph{max entropy} is 
\begin{align}
   \label{eq:HMax_Def}
   H_\Max ( \sigma )  
   & :=  H_{ 1 / 2 }  ( \sigma ) 
   =  \log \left( || \sqrt{ \sigma }  ||_1^2  \right)  \, .
\end{align}
The Schatten 1-norm is denoted by $|| . ||_1$.
The general Schatten $p$-norm of
a Hermitian operator
$\sigma  =  \sum_j  s_j  \ketbra{ s_j }{ s_j }$ is
\begin{align}
   \label{eq:Schatten_Def}
   || \sigma ||_p  
   =  \left[  \Tr  \left(  \sigma^p  \right)  \right]^{1 / p}
   =  \left(  \sum_j  | s_j |^p  \right)^{ 1 / p }  \, ,
\end{align}
for $p  \geq  1$~\cite{Bhatia_97_Matrix}.
$H_\Max$ reflects the discrepancy 
between 
$\sigma$ and the maximally mixed state~\cite[p.~60]{Tomamichel_12_Framework}:
The fidelity between normalized states
$\sigma$ and $\gamma$ is\footnote{
Tomamichel uses the generalized fidelity.
When we evaluate the generalized fidelity,
at least one argument is normalized.
The generalized fidelity therefore simplifies to
the fidelity~\cite[p.~48]{Tomamichel_12_Framework}.}
$F( \sigma, \gamma )  
:=  || \sqrt{ \sigma }  \sqrt{ \gamma }  ||_1$.
$H_\Max$ depends on the fidelity through:
$H_\Max ( \sigma )
=  \log \LParen \Dim \,
    [ F ( \sigma ,  \id / \Dim ) ]^2  \RParen$.

We notate the detector state's R\'enyi entropies as
\begin{align}
   H_\alpha \LParen V  \W(t)  \RParen_\rho
   :=  H_\alpha  (  \rho_\Fwd  )  \, ,
\end{align}
following~\cite{Tomamichel_12_Framework}.
We have now introduced the forward-POVM entropies.
The two-detector state $\rho_\Rev$,
and the entropy $H_\alpha \LParen  \W(t)  V  \RParen$,
are defined analogously.


$H_\Max$ and $H_\Min$, like $H_\vN$,
quantify rates at which 
information-processing and thermodynamic tasks can be performed.
Applications include quantum key distribution, 
randomness extraction, erasure, work extraction, 
and work expenditure
(e.g.,~\cite{Renner_05_Security,delRio_11_Thermodynamic,Tomamichel_12_Framework,Berta_13_Quantum,Tomamichel_16_Quantum,Leditzky_16_Relative,Leditzky_16_Strong,Faist_18_Fundamental}).
Quantum states desired for such tasks cannot be prepared exactly.
\emph{Smoothing} introduces an error tolerance
$\varepsilon  \in  [0, 1)$
into the entropies~\cite{Renner_05_Security}.
Our uncertainty relations for scrambling 
generalize to smooth entropies.
We focus on nonsmooth entropies for simplicity.

\subsection{Entropic uncertainty relations for QI scrambling}
\label{sec:Main_thm_and_analysis}

We can now reconcile the two notions
of quantum operator disagreement,
entropic uncertainty relations of pure QI theory
and information scrambling of high-energy and condensed-matter theory.

%
%
\begin{theorem}
\label{theorem:OTOC_thm_Toma}

The forward and reverse POVMs
satisfy entropic uncertainty relations for scrambling,
\begin{align}
   \label{eq:KP_OTOC}
   & H_\vN  \LParen  V  \W(t)  \RParen_\rho
   +  H_\vN  \LParen  \W(t)  V  \RParen_\rho
   \geq  f ( v_1, v_2 ) 
   \; \text{and} \; \\
   \label{eq:Rastegin_OTOC}
   & H_\alpha  \LParen  V  \W(t)  \RParen_\rho
   +  H_\beta  \LParen  \W(t)  V  \RParen_\rho
   \geq  f ( v_1, v_2 ) \, ,
   %
\end{align}
for 
$\frac{1}{\alpha}  +  \frac{1}{\beta} =  2$.
The bound depends on the OTOC quasiprobability:
\begin{align}
   \label{eq:Full_Bound}
   f ( v_1, v_2 )  
   &  :=  \min_{ j_1,  j_2,  w_1,  w_2 }  \big\{ 
   C_0
   + {\rm Re} \left(  \gV{j_1}  \right)  C_1
   + {\rm Re}  \left(  \gV{j_2}  \right)  C'_1 
   \nonumber \\ & \qquad
   +  {\rm Re}  \left(  \gV{j_1}  \gV{j_2}  
                        \SumKD{\id} ( v_1, w_1, v_2, w_2 )  \right)      C_2
   \\ & \qquad \nonumber
   +  \Verts{  \gV{j_1}  }^2
       \SumKD{\id} ( v_1, w_1, v_1, w_2 )       C'_2
   \\ \nonumber & \qquad
   +  \Verts{  \gV{j_2}  }^2
       \SumKD{\id} ( v_2, w_1, v_2, w_2 )     C''_2
   +  O (g^2)   \Big\} \, .
\end{align}
The real numbers $C$,
and the rest of the $\sim g^2$ terms,
depend essentially on classical probabilities.
Their forms are given below.
\end{theorem}
\noindent 
The $j$ and $w$ dependences of the $C$'s
have been suppressed for conciseness.
Inequality~\eqref{eq:Rastegin_OTOC} can be smoothed
when $(\alpha, \beta)  =  (\infty, 1/2)$.


The uncertainty relations are proved in App.~\ref{section:Proof_MainThm}.
They follow from three general uncertainty relations:
Result~7 in~\cite{Tomamichel_12_Framework},
Corollary~2.6 in~\cite{Krishna_01_Entropic}, and
Ineq.~(13) in~\cite{Rastegin_08_Uncertainty}.
The OTOC POVMs~\eqref{eq:Fwd_POVM} and~\eqref{eq:Rev_POVM}
are substituted into the general uncertainty relations.
The POVMs' maximum overlap, $c$, cannot obviously be inferred 
from parameters chosen, or from measurements taken,
in an OTOC-inference experiment.
We therefore bound $c$, using $\SumKD{\rho}$ and
the Schatten $p$-norm's monotonicity in $p$:
\begin{align}
   \label{eq:Tr_In_Proof}
   - & \log c  \geq
   \log  \Bigg(  \min_{ j_1, j_2, w_1, w_2 }  \Bigg\{
   \\ \nonumber &
   \Tr  \left(     \ProjWt{w_2}  \KrausV{v_2}{j_2}  
   \left[  \KrausV{v_1}{j_1}  \right]^\dag
   \ProjWt{w_1}
   \KrausV{v_1}{j_1}
   \left[  \KrausV{v_2}{j_2}  \right]^\dag  
   \right)  \Bigg\}  \Bigg)  \, .
\end{align}
We substitute in for the $K$'s
from Eq.~\eqref{eq:Kraus}, then multiply out.
In each of several terms,
two $K$'s contribute $\ProjV{v_\ell}$'s,
while two $K$'s contribute $\id$'s.
These terms contain quasiprobaiblity values $\SumKD{\id}$.
We isolate the terms by Taylor-expanding 
the logarithm in the $g$'s.

%
%
\subsection{Analysis}
\label{sec:Analysis_MainThm}

Four points merit analysis:
the POVMs' implications for the butterfly effect,
the form of the bound $f ( v_1, v_2 )$,
simple limits, and
conditions that render the bound nontrivial.

\textbf{Implications for the butterfly effect:}
The weak measurements strengthen an analogy between the OTOC
and the butterfly effect of classical chaos~\cite{Shenker_Stanford_14_BHs_and_butterfly,Roberts_16_Lieb,Aleiner_16_Microscopic,Campisi_16_Thermodynamics}.
In the classical butterfly effect,
a tiny perturbation snowballs into a drastic change.
This perturbation has been likened to
operation by a unitary $V$, in Eq.~\eqref{eq:Overlap_OTOC}.
$V$ should be associated with a weak measurement,
Theorem~\ref{theorem:OTOC_thm_Toma} clarifies.
The measurement is perturbative in $\gV{j_\ell}$.

\textbf{Form of the uncertainty bound $f ( v_1, v_2 )$
for scrambling:}
The bound~\eqref{eq:Full_Bound} contains three terms 
dependent on the quasiprobability $\SumKD{\id}$.
These terms' proportionality to $g^2$
accords with intuition:
Scrambling is a subtle feature of quantum equilibration,
detectable in just many-point correlators.
Likewise, the OTOC quasiprobability governs
high-order terms in the uncertainty bound.
As anticipated in Sec.~\ref{sec_Intuit},
the quasiprobability $\SumKD{\id}$ is evaluated on
the identity operator.
The bound highlights the operator disagreement
without pollution by any state $\rho$.

The quasiprobability-free terms in~\eqref{eq:Full_Bound} 
are ``background terms'':
They contain classical probabilities,
accessible without weak measurements.
The $g$-independent term, 
\begin{align}
   \label{eq_C0}
   C_0
   :=  - \log  \left(  \pV{ j_1 }  \pV{ j_2 }   \, 
                        \Tr  \left(  \ProjW{w_2}  
                        \delta_{ w_1 w_2 }   \right)  \right) \, ,
\end{align}
dominates $f ( v_1, v_2 )$.
The Kronecker delta is denoted by $\delta_{ w_1 w_2 }$.
The two linear terms,
\begin{align}
   & \label{eq:Full_Bound_line2}
   C_{1}  :=
   \frac{ - 2 }{  \ln 2  } \:
   p( v_1 | w_2 )   \:
   \frac{ \Real  \left(  \gV{ j_1 }  \right)  }{ 
             \sqrt{  \pV{ j_1 }  }  }
    \quad \text{and} \nonumber \\ &
    C'_{1}  :=
    \frac{ - 2 }{  \ln 2  }  \:
    p(v_2 | w_2 )   \:
    \frac{ \Real  \left(  \gV{ j_2 }  \right)  }{ 
                \sqrt{ \pV{ j_2 } } }  \, ,
\end{align}
depend on projectors $\Pi$
only through classical probabilities
$p (v_\ell | w_m) 
=  \Tr  \left(  \ProjV{v_\ell}  \ProjWt{w_m}  \right) / 
\Tr \left( \ProjWt{w_m} \right)$.
This $p (v_\ell | w_m)$ equals the conditional probability that,
if the system begins maximally mixed
over the $w_m$ eigenspace of $\W(t)$,
if $V$ is measured, 
outcome $v_\ell$ will obtain.
Such classical dependence characterizes also
the $g^2$ terms suppressed in Eq.~\eqref{eq:Full_Bound},
\begin{align}
   \label{eq:Full_Bound_line4} &
   \frac{- 1}{ \ln 2  }  \:
   p(v_1 | w_2 )  \:
   \Real \left(  \gV{j_1}  \left[  \gV{ j_2 }  \right]^*  \right)
                       \delta_{ v_1 v_2 }  
   \\ \nonumber & 
   +  \frac{ 2 }{  \Tr \left( \ProjW{w_2}  \right)  } \Bigg[ 
      \frac{ \Real  \left(  \gV{ j_1 }  \right)  }{ 
             \sqrt{  \pV{ j_1 }  }  }  \:
    p(v_1 | w_2 ) 
    + \frac{ \Real  \left(  \gV{ j_2 }  \right)  }{ 
                \sqrt{ \pV{ j_2 } } }  \:
    p(v_2 | w_2 )  \:   \Bigg]^2  \, .
\end{align}

The dominance of $C_0$, the $\delta_{w_1 w_2}$ in $C_0$,
and the min ensure that
$w_1 = w_2$ throughout the min's argument.
The first $\SumKD{\id}$ has four arguments,
$( v_1, w_1, v_2, w_2 )$,
constrained only by the $\delta_{ w_1 w_2 }$.
In each other $\SumKD{\id}$, 
the first argument must equal the third,
even before the minimization is imposed.
For example, the second quasiprobability value has the form 
$\SumKD{\id} ( v_1, w_1, v_1, w_2 )$.
The $V$ eigenvalues equal each other,
due to Ineq.~\eqref{eq:Tr_In_Proof}.
One $v_1$ comes from the $\left(  \KrausV{v_1}{j_1}  \right)^\dag$,
and one, from the $\KrausV{v_1}{j_1}$.

\textbf{Nontriviality conditions:}
The R\'enyi entropies are nonnegative:
$H_\alpha ( \sigma )  \geq  0$.
Hence the bound is nontrivial when positive:
$f ( v_1, v_2 )  >  0$.
When the coupling is weak,
the bound is positive when its first term is positive.
The first term simplifies to 
$\min_{ j_1, j_2,  w_2 } \Set{ 
- \log  \left(  \pV{j_1}  \pV{j_2}  \:
\Tr \left( \ProjW{w_2}  \right)  \right)  }$.
The trace is large in the system size, 
equaling $2^{ \Sites - 1 }$ in the spin-chain example.
One might worry that this trace swells the log,
drawing the bound far below zero.

The probabilities $\pV{j_\ell}$ can offset the enormity.
Let us focus on the spin-chain example
and approximate $\pV{j_1}  \approx  \pV{j_2}  \equiv  \pV{j_\ell}$.
Nonnegativity of the log term becomes equivalent to
$\left(  \pV{j_\ell}  \right)^2  \:  2^{ \Sites - 1 }  \leq  1$,
or
$\pV{j_\ell}  \leq  \frac{1}{  2^{ (\Sites - 1) / 2 }  }$.
Strongly measuring a weak-measurement detector
must yield one of $\geq 2^{ (\Sites - 1) / 2 }$ possible outcomes.

Weak measurements as in~\cite{Aharonov_88_How}
satisfy this requirement.
Let each detector manifest as a particle,
e.g., in a potential that defines a dial.
Let $O$ denote the strongly measured detector observable
(e.g., the position $\hat{x}$).
Let $\tilde{O}$ denote the conjugate observable
(e.g., the momentum $\hat{p}$):
$\left[  O ,  \tilde{O}  \right]
=  \pm  i  \hbar$.
Let be prepared in a Gaussian state that peaks sharply at
some $\tilde{O}$ eigenvalue
(e.g., a sharp momentum-space wave packet).
The probabilities $\pV{ j_\ell }$ can be small enough
that $f( v_1, v_2 ) > 0$.
We present an example in Sec.~\ref{sec:Numerics}.

The $g$-free log encodes randomness in
a measurement of a detector that has never coupled
to the system.
Hence the log fails to reflect
disagreement between $V$ and $\W(t)$.
The disagreement manifests in
the $g$-dependent terms.

%
%
\textbf{Simple limits:}
Three simple limits illuminate the bound's behavior:
early times ($t \approx 0$), late times ($t \geq t_*$),
and the weak limit ($g \to 0$).
We focus on a chaotic spin chain, for concreteness.
Numerical simulations (Sec.~\ref{sec:Numerics}) 
support these arguments.

%
%
%
\emph{Early times} ($t  \approx  0$):
$V$ and $\W(t)  \approx  \W$ nontrivially transform just far-apart subsystems.
Hence $\Tr \left(  \ProjWt{w_\ell}  \ProjV{v_m}  \right)
\approx  2^{ \Sites - 2 }$.
Also, $[V,  \W(t) ]  \approx  0$,
so the projectors nearly commute. Hence
$\Tr  \left(  \ProjWt{ w_\ell }  \ProjV{ v_m }  
\ProjWt{ w_{\ell'} }  \ProjV{ v_{m'} }   \right)
\approx  2^{ \Sites  -  2 }  \;
\delta_{ w_\ell  w_{\ell'} }  \delta_{ v_m  v_{m'} }$.
These traces are large, dragging
the $\sim g$ terms in Eq.~\eqref{eq:Full_Bound}, 
and the negative term in~\eqref{eq:Full_Bound_line4}, below zero.
The $g$'s mitigate the dragging's magnitude.
Still, the bound is expected to be relatively loose before $t_*$.


%
%
%
\emph{Late times} ($t  \geq  t_*$):
$V$ can fail to commute with $\W(t)$.
Traces $\Tr  \left(  \ProjWt{ w_\ell }  \ProjV{ v_m }  \ldots  \right)$
will shrink:
Consider a one-qubit system, as a simple illustration.
Suppose that $V = \sigma^z$ and that $\W = \sigma^x$.
Each $\ProjWt{ w_\ell }  \ProjV{ v_m }$
translates roughly into a
$|  \braket{  x_\ell  }{  z_m }  |^2
=  \frac{1}{ \Dim }$.
The traces' smallness tightens the uncertainty bound,
as expected when the system is scrambled
(as explained in the introduction).\footnote{
This expectation is borne out when $v_1 = v_2$, 
as implied by (i) App.~\ref{sec:Why_v1v2} 
and (ii) reasoning, similar to that in the appendix, 
about the $\Verts{ \gV{j} }^2$ terms in Eq.~\eqref{eq:Full_Bound}.
Appendix~\ref{sec:Why_v1v2} 
also shows why the quasiprobability tightens the bound
when (i) $v_1 = - v_2$ and 
(ii) $\gV{j_1} \gV{j_2}$ approximately equals a negative real number.}
The bound likely does not remain at its maximum possible value
at all $t > t_*$, however.
As $\W(t)$ evolves, the bound should fluctuate around 
a relatively large value.

%
%
%
\emph{Weak limit} ($g  \to  0$):
The system fails to couple to the detectors.
The bound~\eqref{eq:Full_Bound} reduces to
$\min_{ w_2 }  \Set{  - \log  \left(  
\pV{j_1}  \pV{j_2}
\Tr \left( \ProjW{w_2}  \right)  \right)  }$.
The probability distribution $\Set{  \pV{j_\ell}  }$
has a spread quantified by the Shannon entropy
$H_\Sh  \left(  \Set{  \pV{j_\ell}  }  \right)
:=  - \sum_{j_\ell}  \pV{j_\ell}  \log  \pV{j_\ell}$. 
The left-hand side of Ineq.~\eqref{eq:KP_OTOC} reduces to
$2 \left[  H_\vN \LParen \W(t)  \RParen_\rho
       +  H_\Sh  \left(  \Set{  \pV{j_\ell}  }  \right)  \right]$.


%
%
\section{Numerical simulations of a spin chain}
\label{sec:Numerics}

We illustrate Theorem~\ref{theorem:OTOC_thm_Toma} with
an interacting spin chain.
The setup and weak-measurement implementation
are described in Sec.~\ref{sec:Spin_Chain_Set_up}.
The detector probabilities $\pV{j_\ell}$,
the weak-measurement Kraus operators $\KrausV{v_\ell}{j_\ell}$,
the couplings $\gV{j_\ell}$, 
and the entropies $H_\alpha$
are presented in Sec.~\ref{sec:Spin_Chain_Calcs}
and calculated in App.~\ref{sec_Spin_Chain_Analytics}.
We present and analyze results 
in Sec.~\ref{sec:Spin_Chain_Results}.

\subsection{Spin-chain setup}
\label{sec:Spin_Chain_Set_up}

Consider a one-dimensional (1D) chain of $\Sites = 8$ qubits.
The OTOC operators manifest as single-qubit Pauli operators:
$V = \sigma^z_1$, and $\W = \sigma^z_\Sites$.
The operators' precise forms do not impact 
our chaotic-system results, however.

%
%
%
\textbf{Model:}
The chain evolves under the power-law quantum Ising Hamiltonian
\begin{align}
   \label{eq:PQIM}
   H_{\rm PQIM}  
   = - J  \sum_{ \ell = 1 }^{ \ell_0 }  \sum_{ j = 1 }^{ \Sites - \ell }
   \frac{1}{ \ell^\zeta }
   \sigma^z_j  \sigma^z_{ j + \ell }
   - h^x  \sum_{ j = 1 }^\Sites   \sigma^x_j
   - \sum_{j = 1}^\Sites  h^z_j  \sigma^z_j
\end{align}
\cite{Swingle_18_Resilience}
(see~\cite{Chen_17_SubsystemDiff} for a similar model).
Each spin $j$ interacts with each spin
that lies within a distance $\ell_0$.
The interaction strength declines with distance 
as a power law controlled by $\zeta > 0$.
We choose $J = 1$, $\zeta = 6$, and $\ell_0 = 5$,
as in~\cite{Swingle_18_Resilience}.
Planck's constant is set to one: $\hbar = 1$.
We set the transverse field $h^x$ to $1.05$.
The longitudinal field $h^z_j  =  0.375 ( -1 )^j$
flips from site to site.

The transverse-field Ising model with a longitudinal field
reproduces our results' qualitative features.
But the power-law quantum Ising model mimics 
all-to-all interactions, such as in the SYK model~\cite{Sachdev_93_Gapless,Kitaev_15_Simple,Polchinski_16_Spectrum,Maldacena_16_Remarks}.
Around $t = t_*$, therefore,
the OTOC decays almost exponentially.
Exponential decay evokes classical chaos,
as discussed in the introduction.

\textbf{Weak-measurement implementation:}
Section~\ref{sec:Analysis_MainThm} guides our implementation,
which parallels~\cite{Aharonov_88_How}.
We illustrate with the forward-protocol weak measurement,
temporarily reinstating operators' hats.

The detector consists of a particle 
that scatters off the system.
The detector could manifest as a photon,
as in circuit QED~\cite{deLange_14_Reversing}
and in purely photonic experiments~\cite{Lundeen_11_Direct}.
Let $\mathbf{\hat{y}}$ denote the longitudinal direction,
which points from the detector's initial position to the system.

Let $\mathbf{\hat{x}}$ denote a transversal direction;
and $\ket{ D }$, the $\mathbf{\hat{x}}$ component of the detector's initial state.
$\ket{ D }$ consists of a Gaussian,
\begin{align}
   \label{eq:Det_State}
   \ket{ D }  =  \frac{1}{ \pi^{ 1 / 4 }  \, \sqrt{ \Delta } }  \:
   \int_{ - \infty }^\infty  dp  \;
   e^{ - p^2 / 2 \Delta^2 }  \:  \ket{ p }  \, ,
\end{align}
centered on the transverse-momentum eigenvalue 
$p \equiv  p_x  = 0$.
$\Delta$ denotes the Gaussian's standard deviation.

The displaced detector position $\hat{x}  -  x_0  \hat{\id}$
couples\footnote{
$\ProjVH{v_\ell}$ can effectively be measured weakly
via coupling of the detector to $V  =  \hat{\sigma}_\ell^z$.
The interaction unitary will have the form
$\exp  \left(  - \frac{i}{ \hbar }  \,  \tilde{g}  
\left[  \hat{x}  \otimes  \hat{\sigma}_\ell^z  \right]  \right)$.
The Pauli operator decomposes as
$\hat{\sigma}_\ell^z  =  \pm  
\left(  2  \ProjVH{ \pm }  -  \hat{\id}  \right)$.
Hence the interaction unitary has the form
$\exp   \left(  \pm  \frac{i}{ \hbar }  \,  \tilde{g}  
                             \left[  \hat{x}  \otimes  \hat{\id}  \right]  \right)
   \exp  \left(  \mp  \frac{2 i }{ \hbar }  \,  \tilde{g}  
                     \left[  \hat{x}  \otimes  \ProjVH{ \pm }  \right]  \right)  \, .$
The Kraus operator becomes 
$\bra{ x_\ell } 
  \exp   \left(  \pm \frac{i}{ \hbar }  \,  \tilde{g}  
                                \left[  \hat{x}  \otimes  \hat{\id}  \right]  \right)
   \exp  \left(  \mp  \frac{2 i }{ \hbar }  \,  \tilde{g}  
                     \left[  \hat{x}  \otimes  \ProjVH{ \pm }  \right]  \right)
   \ket{ D }$.
The lefthand exponential can be absorbed into 
the strong measurement of the detector:
Consider wishing to measure $\ProjVH{ + }$ weakly.
Instead of measuring the detector's 
$\Set{ \ket{x_\ell} }$ strongly,
one measures $\Set{ 
e^{ - \frac{i}{ \hbar }  \tilde{g}  \hat{x}  }
\ket{ x_\ell }  }$.}
to the system's $\ProjVH{v_\ell}$.
[The displacement prevents the minimization in~\eqref{eq:Full_Bound}
from choosing the detector-measurement outcome $x = 0$.
This choice would set $\gV{x_\ell}$ to $\gV{x_0} = 0$,
eliminating the weak measurement.]
The interaction unitary has the form
\begin{align}
   \hat{V}_\inter  
   \label{eq:Ex_V_inter_Exp}
   & =  \exp  \left(   - \frac{i}{ \hbar }  \,  \tilde{g}  
   \left[  \hat{x}  -  x_0  \hat{\id}  \right]  
   \otimes  \ProjVH{v_\ell}  \right)  \\
   \label{eq:Ex_V_inter}
   & =  \hat{\id}  
   +  \left(  e^{ - \frac{i}{ \hbar }  \tilde{g}  
                \left(  \hat{x}  -  x_0  \hat{\id}  \right)  }
   - \hat{\id}  \right)  \otimes  \ProjVH{v_\ell}  \, .
\end{align}
The interaction strength $\tilde{g}$ governs 
the outcome-dependent coupling $\gVH{j_\ell}$.
Numerical experiments show that
$\tilde{g} = 0.02$ and $x_0 = 10$
keep $\frac{  \gV{j_\ell}  }{ \sqrt{ \pV{x_\ell} } }$ 
perturbatively small while strengthening the bound.

The detector's $\hat{x}$  
is measured strongly.
Let $L > 0$ denote the measurement's precision.
Positions $x_1$ and $x_2$ can be distinguished
if they lie a distance $| x_2 - x_1 |  \geq  L$ apart.
Hence the classical register has a discrete spectrum $\Set{ x_\ell }$.
We simulated a register whose $L = 0.1$.

\subsection{Analytical ingredients in
spin-chain uncertainty relation}
\label{sec:Spin_Chain_Calcs}

Analytical results are presented here:
the detector probability 
$\pVH{j_\ell}  \equiv  \pVH{x_\ell}$,
the weak-measurement Kraus operators 
$\KrausVH{v_\ell}{j_\ell}  \equiv  \KrausVH{v_\ell}{x_\ell}$,
the coupling strengths 
$\gVH{j_\ell}  \equiv  \gVH{x_\ell}$,
and the entropies $H_\alpha$.
We derive these results and check their practicality
in App.~\ref{sec_Spin_Chain_Analytics}.

Consider preparing the detector in $\ket{D}$,
then measuring $\hat{x}$.
The measurement has a probability
$\pVH{x_\ell}  L  =  |  \braket{ x_\ell }{ D }  |^2  L$
of yielding a position within $L$ of  $x_\ell$.
By Eq.~\eqref{eq:Det_State},
\begin{align}
   \label{eq:Ex_Prob}  
   \pVH{x_\ell}  L  
   =  \frac{ L  \Delta }{ \sqrt{\pi}  \,  \hbar  }  \:
   e^{ - \Delta^2  ( x_\ell )^2  /  \hbar^2  }  \, .
\end{align}
The weak-measurement Kraus operators have the form
\begin{align}
   \KrausVH{ v_\ell }{ x_\ell }
   \label{eq:Ex_Kraus_Help5}
   & =   \sqrt{ \pVH{x_\ell}  } \:  \hat{\id}  
   +  \gVH{x_\ell}   \ProjVH{v_\ell} \, .
\end{align}
The outcome-dependent coupling is
\begin{align}
   \label{eq:Ex_gV}  
   \gVH{x_\ell}
   =  \sqrt{ \pVH{x_\ell}  } 
       \left(  e^{ - \frac{i}{ \hbar }  \,  \tilde{g}  
                (  x_\ell  -  x_0  ) }
                - 1  \right)   \, .
\end{align}
The R\'enyi-$\alpha$ entropy 
limits, as $\alpha \to \infty$, to
\begin{align}
   H_\Min  \LParen  \hat{V}  \hat{\W}(t)  \RParen_{ \hat{\rho} }
   & =  H_\Min  \Big( \Big\{
   \pVH{ j_1 }
   \\ & \quad
   +  2 \sqrt{  \pVH{ j_1 }  }  \:
   \Real  \left(  \gVH{ j_1 }  
                    \Tr  \left(  \ProjWtH{ w_1}  \ProjVH{ v_1 }  \hat{\rho}  \right)  \right)
   \nonumber \\ \nonumber &  \quad 
   +  \Verts{  \gVH{ j_1 }  }^2
   \Tr  \left(  \ProjVH{ v_1 }  \ProjWtH{ w_1 }     
                  \ProjVH{ v_1 }  \hat{\rho}  \right)
   \Big\}_{ v_1, j_1, w_1 }  \Big)  \, .
\end{align}
The other entropies have analogous forms.

\textbf{Entropies $H_\alpha$:}
Let us remove operators' hats.
We illustrate the entropies' analytical forms with 
\begin{align}
   & H_\Min  \LParen  V  \W(t)  \RParen_\rho
   \equiv  H_\Min ( \rho_\Fwd )  \\
   \label{eq:Ex_HMin}
   & =  H_\Min  \left(  \Set{
   \Tr  \left(  \KrausF{v_1}{ j_1 }{ w_1 }^\dag
                   \KrausF{v_1}{ j_1 }{ w_1 }  \:
                   \rho  \right)
   }_{ v_1, j_1, w_1 }  \right)  \, .
\end{align}
The measurement operators have the form
\begin{align}
   & \KrausF{v_1}{ j_1 }{ w_1 }^\dag
                   \KrausF{v_1}{ j_1 }{ w_1 }
   =  \left(  \KrausV{v_1}{ j_1 }  \right)^\dag
   \ProjWt{ w_1 }      \KrausV{v_1}{ j_1 } \, ,
\end{align}
by Eq.~\eqref{eq:Fwd_POVM}.
We substitute in from Eq.~\eqref{eq:Kraus},
multiply out, and substitute into Eq.~\eqref{eq:Ex_HMin}:
\begin{align}
   H_\Min  \LParen  V  \W(t)  \RParen_\rho
   & =  H_\Min  \Big( \Big\{
   \pV{ j_1 }
   \\ & \quad
   +  2 \sqrt{  \pV{ j_1 }  }  \:
   \Real  \left(  \gV{ j_1 }  
                    \Tr  \left(  \ProjWt{ w_1}  \ProjV{ v_1 }  \rho  \right)  \right)
   \nonumber \\ \nonumber &  \quad 
   +  \Verts{  \gV{ j_1 }  }^2
   \Tr  \left(  \ProjV{ v_1 }  \ProjWt{ w_1 }     
                  \ProjV{ v_1 }  \rho  \right)
   \Big\}_{ v_1, j_1, w_1 }  \Big)  \, .
\end{align}
The other entropies have analogous forms.

\subsection{Spin-chain results}
\label{sec:Spin_Chain_Results}

\begin{figure}[hbt]
\centering
\def\svgwidth{.49\textwidth}
\begingroup
  \setlength{\unitlength}{\svgwidth}
  \providecommand\rotatebox[2]{#2}%
  \begin{picture}(1,0.66666667)
    \setlength\tabcolsep{0pt}
    \put(0,0){\includegraphics[width=\unitlength]{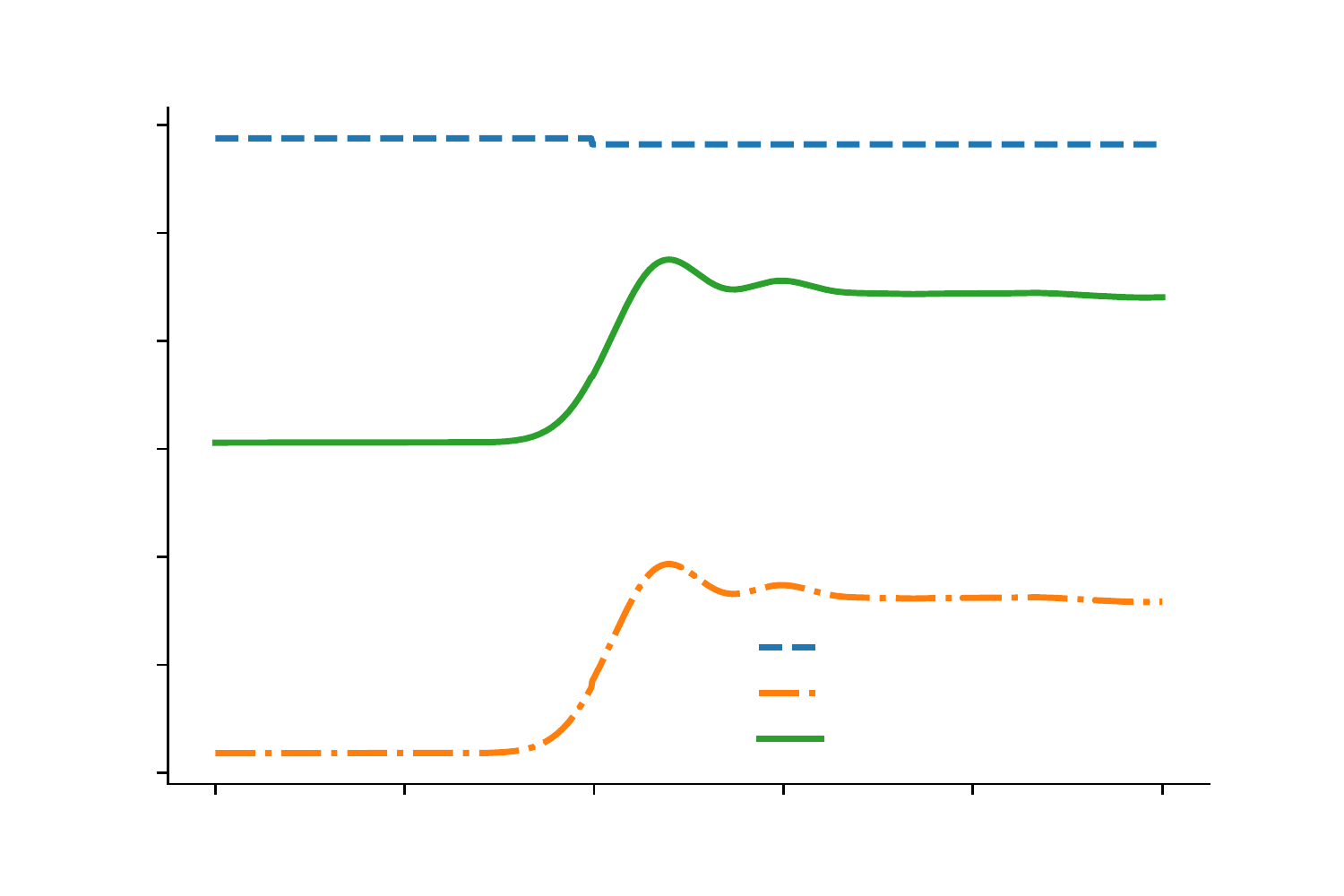}}
    \put(0.15286688,0.0445515){\makebox(0,0)[lt]{\smash{0}}}
    \put(0.29377596,0.0445515){\makebox(0,0)[lt]{\smash{2}}}
    \put(0.43468506,0.0445515){\makebox(0,0)[lt]{\smash{4}}}
    \put(0.57559415,0.0445515){\makebox(0,0)[lt]{\smash{6}}}
    \put(0.71650322,0.0445515){\makebox(0,0)[lt]{\smash{8}}}
    \put(0.85005192,0.0445515){\makebox(0,0)[lt]{\smash{10}}}
    \put(0.4082248,0.00590365){\makebox(0,0)[lt]{\smash{Time (units of $1/J$)}}}
    \put(0.02826454,0.08286997){\makebox(0,0)[lt]{\smash{-0.06}}}
    \put(0.028226454,0.16322321){\makebox(0,0)[lt]{\smash{-0.04}}}
    \put(0.02826454,0.24357645){\makebox(0,0)[lt]{\smash{-0.02}}}
    \put(0.0422555,0.32392969){\makebox(0,0)[lt]{\smash{0.00}}}
    \put(0.0422555,0.40428293){\makebox(0,0)[lt]{\smash{0.02}}}
    \put(0.0422555,0.48463617){\makebox(0,0)[lt]{\smash{0.04}}}
    \put(0.0422555,0.56498941){\makebox(0,0)[lt]{\smash{0.06}}}
    \put(0.01679919,0.31326242){\rotatebox{90}{\makebox(0,0)[lt]{\smash{Bits}}}}
    \put(0.62952834,0.17390248){\makebox(0,0)[lt]{\smash{Terms $\propto g$}}}
    \put(0.62952834,0.13893981){\makebox(0,0)[lt]{\smash{Terms $\propto g^{2}$}}}
    \put(0.62952834,0.10397714){\makebox(0,0)[lt]{\smash{Sum}}}
  \end{picture}
\endgroup
\caption{\caphead{Greatest coupling-dependent contributions to
the entropic uncertainty bound for scrambling:} 
We numerically simulated a one-dimensional chain of $\Sites = 8$ qubits 
evolving under the power-law quantum Ising Hamiltonian~\eqref{eq:PQIM}.
The nearest-neighbor coupling $J = 1$,
the transverse field $h^x  =  1.05$,
and $\zeta = 6$ and $\ell_ 0 = 5$ govern the interactions' power-law decay.
The system was initialized in the Gibbs state
$\rho = e^{ - \beta H } / Z$ at inverse temperature $\beta = 1$.
The weak-coupling strength $\tilde{g} = 0.02$.
The out-of-time-ordered-correlator (OTOC) operators $V$ and $\W$
manifest as single-qubit Pauli operators
localized on opposite sides of the chain:
$V = \sigma^z_1$, and $\W = \sigma^z_\Sites$.
The greatest coupling-dependent contributions to
the entropic uncertainty bound $f( v_1{=}1, v_2{=}{-1} )$ 
[Eq.~\eqref{eq:Full_Bound}] are plotted against time, 
measured in units of $1 / J$.
The bound tightens at the scrambling time $t \approx t_*$.
This growth confirms that 
Theorem~\ref{theorem:OTOC_thm_Toma} unifies
two notions of operator disagreement,
entropic uncertainty relations and information scrambling.}
\label{fig:RHS_Zoom}
\end{figure}
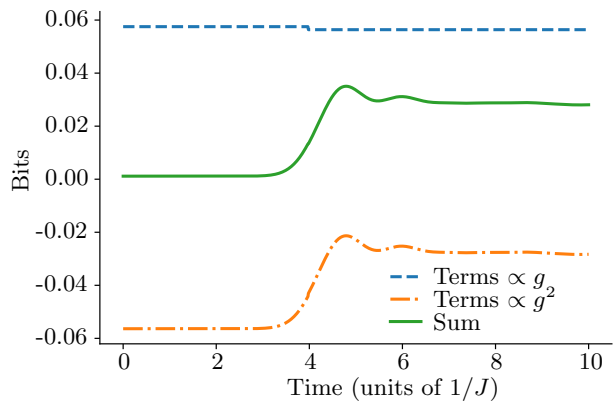

\begin{figure}[hbt]
\centering
\def\svgwidth{.49\textwidth}
\begingroup
  \setlength{\unitlength}{\svgwidth}
  \providecommand\rotatebox[2]{#2}
  \begin{picture}(1,0.66666667)
    \setlength\tabcolsep{0pt}
    \put(0,0){\includegraphics[width=\unitlength]{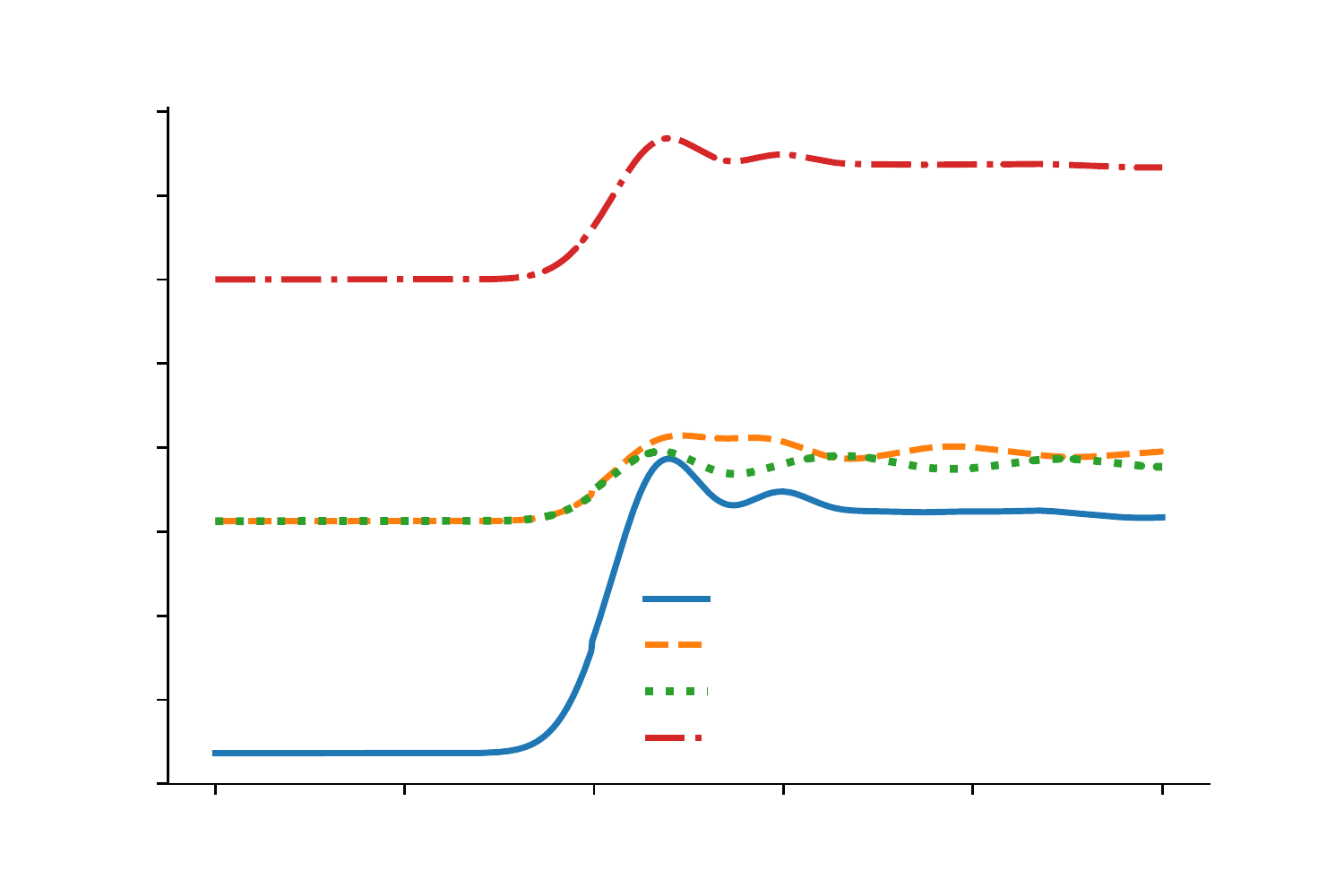}}
    \put(0.15286688,0.0445515){\makebox(0,0)[lt]{\smash{0}}}
    \put(0.29377596,0.0445515){\makebox(0,0)[lt]{\smash{2}}}
    \put(0.43468506,0.0445515){\makebox(0,0)[lt]{\smash{4}}}
    \put(0.57559415,0.0445515){\makebox(0,0)[lt]{\smash{6}}}
    \put(0.71650322,0.0445515){\makebox(0,0)[lt]{\smash{8}}}
    \put(0.85005192,0.0445515){\makebox(0,0)[lt]{\smash{10}}}
    \put(0.4082248,0.00590365){\makebox(0,0)[lt]{\smash{Time (units of $1/J$)}}}
    \put(0.03226454,0.07487807){\makebox(0,0)[lt]{\smash{-0.06}}}
    \put(0.03226454,0.1371179){\makebox(0,0)[lt]{\smash{-0.05}}}
    \put(0.03226454,0.19935774){\makebox(0,0)[lt]{\smash{-0.04}}}
    \put(0.03226454,0.26159759){\makebox(0,0)[lt]{\smash{-0.03}}}
    \put(0.03226454,0.32383742){\makebox(0,0)[lt]{\smash{-0.02}}}
    \put(0.03226454,0.38607726){\makebox(0,0)[lt]{\smash{-0.01}}}
    \put(0.0422555,0.4483171){\makebox(0,0)[lt]{\smash{0.00}}}
    \put(0.0422555,0.51055693){\makebox(0,0)[lt]{\smash{0.01}}}
    \put(0.0422555,0.57279677){\makebox(0,0)[lt]{\smash{0.02}}}
    \put(0.01679919,0.31326242){\rotatebox{90}{\makebox(0,0)[lt]{\smash{Bits}}}}
    \put(0.5434317,0.21081828){\makebox(0,0)[lt]{\smash{All terms $\propto g^{2}$}}}
    \put(0.5434317,0.17685561){\makebox(0,0)[lt]{\smash{Term $\propto \SumKD{\id} ( v_1, w_1, v_1, w_2 )$}}}
    \put(0.5434317,0.14224189){\makebox(0,0)[lt]{\smash{Term $\propto \SumKD{\id} ( v_2, w_1, v_2, w_2 )$}}}   
    \put(0.5434317,0.10762818){\makebox(0,0)[lt]{\smash{Term $\propto \SumKD{\id} ( v_1, w_1, v_2, w_2 )$}}}   
  \end{picture}
\endgroup
\caption{\caphead{Quasiprobability's contribution to
the entropic uncertainty bound for scrambling:} 
The quasiprobability $\SumKD{\id}$ governs 
three terms in the bound 
$f( v_1{=}1, v_2{=}{-1} )$ [Eq.~\eqref{eq:Full_Bound}];
these terms are plotted against time.
The system parameters are 
those described below Fig.~\ref{fig:RHS_Zoom}.}
\label{fig:Quasiprobs}
\end{figure}
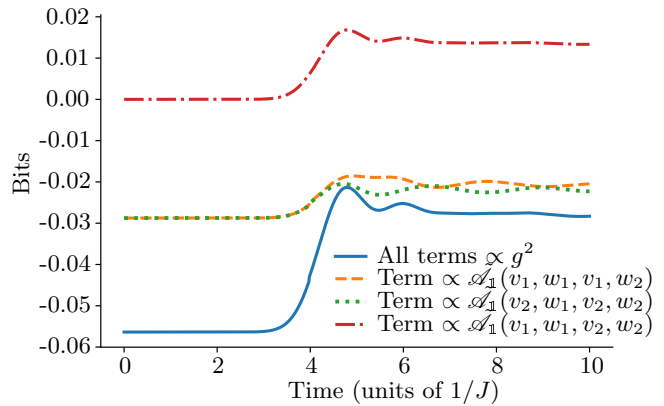

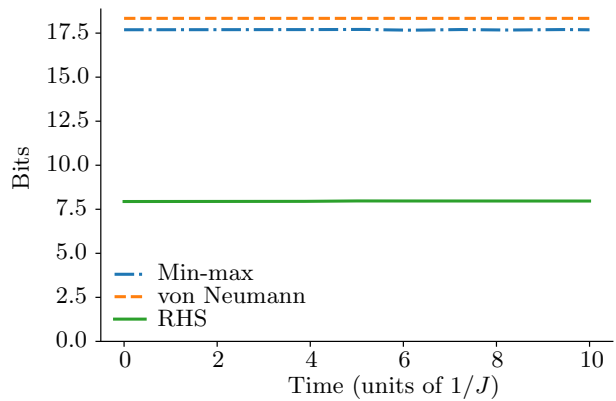
\begin{figure}[hbt]
\centering
\def\svgwidth{.49\textwidth}
\begingroup
  \setlength{\unitlength}{\svgwidth}
  \providecommand\rotatebox[2]{#2}
  \begin{picture}(1,0.66666667)
    \setlength\tabcolsep{0pt}
    \put(0,0){\includegraphics[width=\unitlength]{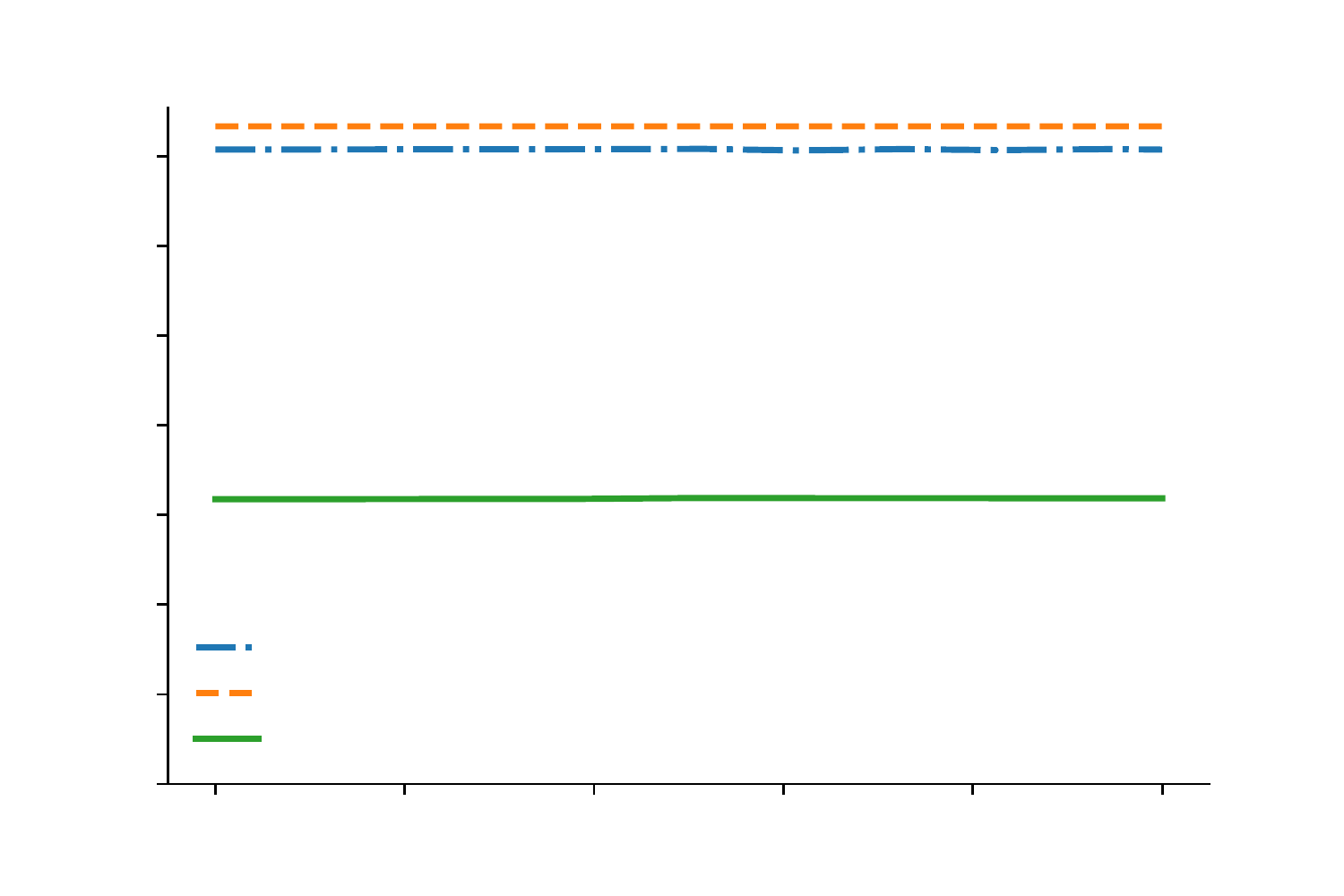}}
    \put(0.15286688,0.0445515){\makebox(0,0)[lt]{\smash{0}}}
    \put(0.29377596,0.0445515){\makebox(0,0)[lt]{\smash{2}}}
    \put(0.43468506,0.0445515){\makebox(0,0)[lt]{\smash{4}}}
    \put(0.57559415,0.0445515){\makebox(0,0)[lt]{\smash{6}}}
    \put(0.71650322,0.0445515){\makebox(0,0)[lt]{\smash{8}}}
    \put(0.85005192,0.0445515){\makebox(0,0)[lt]{\smash{10}}}
    \put(0.4082248,0.00590365){\makebox(0,0)[lt]{\smash{Time (units of $1/J$)}}}
    \put(0.05697627,0.07454427){\makebox(0,0)[lt]{\smash{0.0}}}
    \put(0.05697627,0.14125333){\makebox(0,0)[lt]{\smash{2.5}}}
    \put(0.05697627,0.20796239){\makebox(0,0)[lt]{\smash{5.0}}}
    \put(0.05697627,0.27467145){\makebox(0,0)[lt]{\smash{7.5}}}
    \put(0.0422555,0.34138052){\makebox(0,0)[lt]{\smash{10.0}}}
    \put(0.0422555,0.40808957){\makebox(0,0)[lt]{\smash{12.5}}}
    \put(0.0422555,0.47479865){\makebox(0,0)[lt]{\smash{15.0}}}
    \put(0.0422555,0.54150769){\makebox(0,0)[lt]{\smash{17.5}}}
    \put(0.01618576,0.31326242){\rotatebox{90}{\makebox(0,0)[lt]{\smash{Bits}}}}
    \put(0.21064814,0.17590248){\makebox(0,0)[lt]{\smash{Min-max}}}
    \put(0.21064814,0.14193981){\makebox(0,0)[lt]{\smash{von Neumann}}}
    \put(0.21064814,0.10697714){\makebox(0,0)[lt]{\smash{RHS}}}
  \end{picture}
\endgroup
\caption{\caphead{Left-hand and right-hand sides of
two entropic uncertainty relations for information scrambling:}
The orange, dashed curve illustrates
the $H_\vN + H_\vN$ of Ineq.~\eqref{eq:KP_OTOC}.
The blue, dash-dotted curve illustrates
the $H_\Min + H_\Max$ of Ineq.~\eqref{eq:Rastegin_OTOC}
for $(\alpha, \beta) = (\infty, 1/2)$.
The green, solid curve of Eq.~\eqref{eq:Full_Bound}
illustrates the bound $f ( v_1{=}1, v_2{=}-1 )$.
The system parameters are those described below Fig.~\ref{fig:RHS_Zoom}.
The bound's tightening is undetectable due to the $y$-axis scale.}
\label{fig:LHSs_RHS}
\end{figure}

Figures~\ref{fig:RHS_Zoom}-\ref{fig:LHSs_RHS} illustrate
the entropic uncertainty relations for information scrambling
[Ineqs.~\eqref{eq:KP_OTOC} 
and~\eqref{eq:Rastegin_OTOC}]
in the characteristic parameter regime
detailed in Sec.~\ref{sec:Spin_Chain_Calcs}.
Time is measured in units of the inverse coupling, $1 / J = 1$.
The scrambling time $t_*  \approx  4$,
as reflected by (i) the quasiprobability's sharp change 
in Fig.~\ref{fig:Quasiprobs} and 
(ii) the OTOC's decay in omitted plots.

%
%

%
%

%
%

Figure~\ref{fig:RHS_Zoom} shows 
the greatest time-dependent contributions to 
the bound $f ( v_1, v_2 )$ [Eq.~\eqref{eq:Full_Bound}].
Choosing $v_1  =  - v_2$ tightens the bound
(see App.~\ref{sec:Why_v1v2}),
so we focused on $v_1 = 1$ and $v_2 = -1$.
The bound grows at $t = t_*$, confirming expectations:
At the scrambling time, the OTOC drops.
A decayed OTOC reflects noncommutation of $V$ and $\W(t)$.
The worse two operators commute,
the stronger their entropic uncertainty relations;
the stronger the uncertainty bound $f( v_1, v_2 )$.
Hence Theorem~\ref{theorem:OTOC_thm_Toma} unites
information scrambling and OTOCs with entropic uncertainty relations,
as claimed.

Figure~\ref{fig:Quasiprobs} shows the quasiprobability's contribution
to the uncertainty bound~\eqref{eq:Full_Bound}.
Figure~\ref{fig:LHSs_RHS} shows 
the LHS of Ineq.~\eqref{eq:KP_OTOC} 
($H_\vN + H_\vN$),
the LHS of Ineq.~\eqref{eq:Rastegin_OTOC}
at $(\alpha, \beta) = (\infty, 1/2)$
($H_\Min + H_\Max$),
and the shared RHS $f ( v_1, v_2 )$.  
Figure~\ref{fig:LHSs_RHS} is more zoomed-out than 
Fig.~\ref{fig:Quasiprobs};
hence the tightening is too small to detect.
This reduced visibility is expected: Scrambling is 
a subtle, high-order stage of quantum equilibration.
It manifests in the $g^2$ terms of $f ( v_1, v_2 )$, just as
$\SumKD{\rho}$ can be inferred from
high-order terms in weak-measurement experiments~\cite{NYH_17_Jarzynski,NYH_18_Quasi}.

The LHSs lie $\sim 10$ bits above the bound.
The gap stems from the 
$\Tr \left(  \ProjW{w_2}  \right)  
= 2^{ \Sites - 1 }$
in Eq.~\eqref{eq:Full_Bound}.
This gap bodes ill for the large-system limit, $\Sites \to \infty$,
of interest in holography.
But the gap scales only linearly, not exponentially, with $\Sites$.
Furthermore, small gaps would follow from
many of today's experiments (e.g.,~\cite{Li_16_Measuring}).
Additionally, Sec.~\ref{sec:Beyond} presents
weak-measurement entropic uncertainty relations
independent of scrambling.
Those uncertainty relations need not have such a gap.
We will illustrate with a qubit example
whose bound is tight at zeroth order in $g$,
in Sec.~\ref{sec:Beyond}.

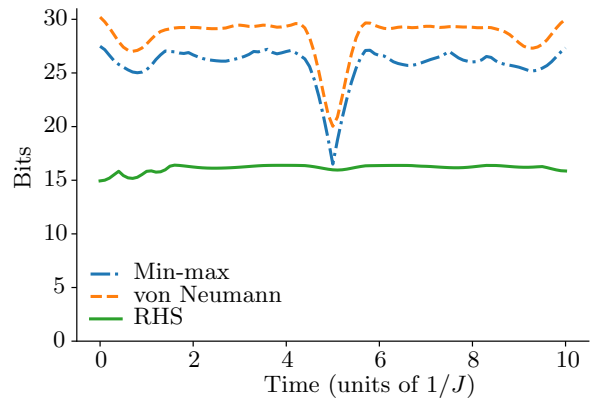
\begin{figure}[hbt]
\centering
\def\svgwidth{.49\textwidth}
\begingroup
  \setlength{\unitlength}{\svgwidth}
  \providecommand\rotatebox[2]{#2}
  \begin{picture}(1,0.66666667)
    \setlength\tabcolsep{0pt}%
    \put(0,0){\includegraphics[width=\unitlength]{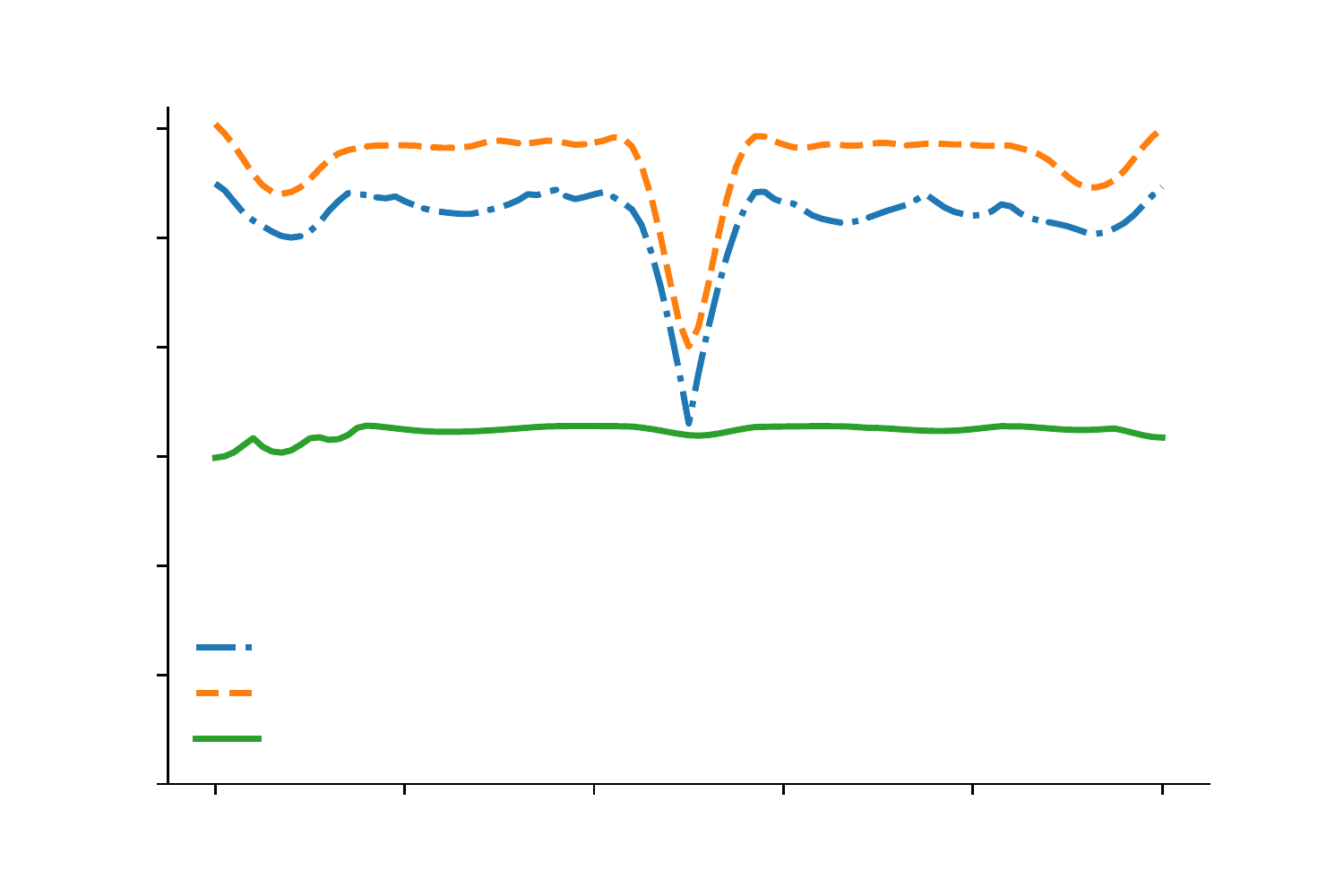}}
    \put(0.15286688,0.0445515){\makebox(0,0)[lt]{\smash{0}}}
    \put(0.29377596,0.0445515){\makebox(0,0)[lt]{\smash{2}}}
    \put(0.43468506,0.0445515){\makebox(0,0)[lt]{\smash{4}}}
    \put(0.57559415,0.0445515){\makebox(0,0)[lt]{\smash{6}}}
    \put(0.71650322,0.0445515){\makebox(0,0)[lt]{\smash{8}}}
    \put(0.85005192,0.0445515){\makebox(0,0)[lt]{\smash{10}}}
    \put(0.4082248,0.00590365){\makebox(0,0)[lt]{\smash{Time (units of $1/J$)}}}
    \put(0.08907552,0.07454427){\makebox(0,0)[lt]{\smash{0}}}
    \put(0.08907552,0.15580117){\makebox(0,0)[lt]{\smash{5}}}
    \put(0.07435474,0.23705807){\makebox(0,0)[lt]{\smash{10}}}
    \put(0.07435474,0.31831499){\makebox(0,0)[lt]{\smash{15}}}
    \put(0.07435474,0.39957189){\makebox(0,0)[lt]{\smash{20}}}
    \put(0.07435474,0.48082878){\makebox(0,0)[lt]{\smash{25}}}
    \put(0.07435474,0.5620857){\makebox(0,0)[lt]{\smash{30}}}
    \put(0.03028501,0.31326242){\rotatebox{90}{\makebox(0,0)[lt]{Bits}}}
    \put(0.21064814,0.17690248){\makebox(0,0)[lt]{\smash{Min-max}}}
    \put(0.21064814,0.14293981){\makebox(0,0)[lt]{\smash{von Neumann}}}
    \put(0.21064814,0.10897714){\makebox(0,0)[lt]{\smash{RHS}}}
  \end{picture}
\endgroup
\caption{\caphead{
Strengthened bound in exceptional parameter regime:}
The system parameters have the values 
below Fig.~\ref{fig:RHS_Zoom}, with three exceptions.
First, the initial state $\rho$ is a $\W(t)$ eigenstate, wherein
the time $t$ is evaluated at the scrambling time $t_*$.
Second, the $\W(t)$ measurements in
the positive-operator-valued measures~\eqref{eq:Fwd_POVM} 
and~\eqref{eq:Rev_POVM} are fine-grained
[are measurements of a $\W(t)$ eigenbasis,
rather than measurements of $\W(t)$].
The orange, dashed curve illustrates $H_\vN + H_\vN$
[Ineq.~\eqref{eq:KP_OTOC}].
The blue, dash-dotted curve illustrates $H_\Min + H_\Max$ 
[Ineq.~\eqref{eq:Rastegin_OTOC} at 
$(\alpha, \beta) = (\infty, 1/2)$].
The upper curves drop to within 0.53 bits of the bound
(the green, solid curve).
Third, the Hamiltonian interaction strength $\tilde{g} = 0.16$,
rendering the measurement-dependent coupling strengths 
$\gV{x_1, x_2}$
comparable to the detector probabilities $\pV{x_1, x_2}$.
This comparability invalidates the Taylor expansion 
that leads to Eq.~\eqref{eq:Full_Bound}.
The bound given by~\eqref{eq:Full_Bound}
appears as the green, solid curve.}
\label{fig:Larger_g}
\end{figure}

Figure~\ref{fig:Larger_g} illustrates how tight the bound can grow
in an exceptional parameter regime.
The top curves represent $H_\Min + H_\Max$ and $H_\vN  +  H_\vN$.
These curves dip at $t \approx t_*$ because
(i) $\rho$ is a $\W(t  \approx  t_*)$ eigenstate and
(ii) the POVMs' $\W(t)$ measurements 
are fine-grained---are replaced with measurements of 
$\Set{  U^\dag \ket{ w_\ell,  \DegenW{w_\ell}  } }$.
The POVM outcomes become highly predictable
around $t_*$, so
the bound grows tight to within 0.53 bits.\footnote{
In addition to choosing $\rho$ and to fine-graining,
we raised the interaction strength to $\tilde{g} = 0.16$.
The outcome-dependent coupling strengths
$\gV{ x_\ell }$ are comparable to 
the detector probabilities:
$\gV{ x_\ell }  \approx  \pV{x_\ell }$.
This comparability invalidates the Taylor expansion
that leads to Eq.~\eqref{eq:Full_Bound}.
Equation~\eqref{eq:Tr_In_C_2} in App.~\ref{section:Proof_MainThm}
gives the pre-Taylor-expansion bound.
This bound appears as the solid, green, bottom curve
in Fig.~\ref{fig:Larger_g}.
The bound would rise more than in the earlier figures,
if the POVMs' $\W(t)$ measurements remained fine-grained:
The large $g$'s would magnify the $\SumKD{\id}$ term's rise.
Since the $\W(t)$ measurements are fine-grained,
the POVMs cease to capture the spirit of scrambling,
defined in terms of local $V$ and $\W$.
Hence we should not necessarily expect scrambling to lift the bound.
}

%
%

Our numerics emphasize the scrambling Hamiltonian $H_{\rm PQIM}$,
which is nonintegrable.
Integrable Hamiltonians' OTOCs revive and decay repeatedly,
as information recollects from across the system and spreads again.
The revivals and decays lift and suppress $f ( v_1, v_2 )$,
we have confirmed using a transverse-field Ising model.
The relevant plots are omitted but appear at~\cite{Data_Avail}.

%
%
%
\section{Extension to higher-point OTOCs}
\label{sec:K_OTOCs}

Higher-point OTOCs reflect later, subtler stages 
of QI scrambling and many-body equilibration.
$F(t)$ has been generalized to
the \emph{$\BarK$-fold OTOC}~\cite{Roberts_16_Chaos,Haehl_17_Classification,Haehl_17_Thermal,NYH_18_Quasi,Dressel_18_Strengthening,Haehl_18_Fine}
\begin{align}
   \label{eq:K_OTOC}
   \OTOCK(t)  & :=  \langle 
   A(t_1)  B (t_2)  C (t_3)  \ldots,  
   E ( t_{ \BarK } )   F ( t_{ \BarK + 1 } )   G ( t_{ \BarK + 2 } )   
   \nonumber \\ & \qquad \times \ldots
   Q ( t_{ 2 \BarK - 1 } )   R ( t_{ 2 \BarK } ) \rangle  \, .
\end{align}
We follow the notation in~\cite{NYH_18_Quasi}.
This $2 \BarK$-point correlator is labeled by
$\BarK  =  1 , 2 , 3 , \ldots$
The conventional OTOC corresponds to $\BarK = 2$.
If $\OTOCK(t)  =  \expval{ \W(t) V  \ldots  \W(t) V }$,
the correlator encodes $\BarK$ 
time reversals, as concretized in 
Schwinger-Keldysh path integrals~\cite{Haehl_17_Classification}
and in the weak-measurement scheme~\cite{NYH_17_Jarzynski,NYH_18_Quasi}.
Higher-point OTOCs $\OTOCK(t)$
equilibrate at later times $t_*^\ParenK  \sim  (\BarK  -  1) t_*$~\cite{Haehl_18_Fine}
and can be inferred from sequences of 
$2 \BarK - 1$ weak measurements.

$\OTOCK(t)$ equals a coarse-graining of
a quasiprobability distribution 
$\SumKD{\rho}^\ParenK$~\cite{NYH_18_Quasi}.
$\SumKD{\id}^\ParenK$ governs 
terms $\propto g^{ 2 ( \BarK - 1 ) }$ in
an entropic uncertainty relation for scrambling.
Denote the eigenvalues of $A(t_1), B(t_2), \ldots$
by $a, b, \ldots$
Denote the eigensubspace projectors by
$\Pi^{ A(t_1) }_{ a }, \Pi^{ B(t_2) }_{ b }, \ldots$
The forward POVM consists of 
a weak measurement of $\Pi^{ R ( t_{ 2 \BarK } ) }_r$,
followed by a weak measurement of $\Pi^{ Q ( t_{ 2 \BarK - 1 } ) }_q$,
and so on, until
a weak measurement of $\Pi^{ G ( t_{ \BarK + 2 } ) }_g$,
followed by a strong measurement of $F ( t_{ \BarK + 1 } )$.
The reverse POVM consists of a strong measurement of $A( t_1 )$,
followed by a weak measurement of $\Pi^{ B ( t_2 ) }_b$,
followed by more weak measurements,
until a weak measurement of $\Pi^{ E ( t_{ \BarK } ) }_e$.

The weak measurement of an observable
$\Theta = B(t_2), C(t_3),  \ldots$
is represented by a Kraus operator 
$\KrausGen{\Theta}{ \theta_\alpha }{j_\alpha}
=  \pGen{\Theta}{ j_\alpha }  \id  
+  \gGen{\Theta}{ j_\alpha }  \Pi^\Theta_\alpha$.
The $j_\alpha$ denotes the weak measurement's outcome,
$\pGen{\Theta}{ j_\alpha }$ denotes the detector probability,
and $\gGen{\Theta}{ j_\alpha }$ denotes
the outcome-dependent weak-coupling strength.

The von Neumann uncertainty relation has the form
\begin{align}
   \label{eq:K_OTOC_Unc_vN}
   & H \LParen  A ( t_1 )  B ( t_2 )  \ldots  E ( t_{ \BarK } )  \RParen
      \\ \nonumber &  \quad
   +  H \LParen  R ( t_{ 2 \BarK } )  Q ( t_{ 2 \BarK - 1 } )   \ldots
                         F ( t_{ \BarK + 1 } )  \RParen
   \nonumber \\ & 
   \geq  - \log  \Big(  \pGen{ B (t_2) }{ j_b }  
                               \pGen{ C (t_3) }{ j_c }  \ldots  
                               \pGen{ E ( t_{ \BarK } ) }{ j_e }
                               \pGen{ G ( t_{ \BarK + 2 } ) }{ j_g }  \ldots
                               \pGen{ Q (t_{ 2 \BarK - 1} ) }{ j_q }  
   \nonumber \\ & \qquad  \qquad  \;  \times
   \Tr  \left(  \Pi^{ A ( t_1 ) }_a  
                   \Pi^{ F ( t_{ \BarK + 1 } ) }_f  \right)
   \Big)  \nonumber \\ & \quad
   + ( g\text{-dependent terms} )  \, .
\end{align}
The term 
\begin{align}
   & \left(  \gGen{B (t_2) }{j_b}  
                \gGen{C (t_3) }{j_c}  \ldots
                \gGen{ E ( t_{ \BarK } ) }{ j_e }  \right)
   \left(  \gGen{ G ( t_{ \BarK + 2 } ) }{ j_g }  \ldots
            \gGen{  Q ( t_{ 2 \BarK - 1 } )  }{ j_q }  \right)
   \nonumber \\ & \quad \times
   \SumKD{\id}^\ParenK   ( r , q , \ldots, a )
\end{align}
contains the quasiprobability behind the $\BarK$-fold OTOC.\footnote{
Entropic uncertainty relations for $\geq 3$ measurements
have been derived~\cite{Wehner_10_Entropic}.
Could such relations contain $\BarK$-fold OTOCs?
The match appears unnatural, for two reasons.
First, consider the minimal generalization of $F(t)$,
in which every observable equals $\W(t)$ or $V$:
$\expval{ \W(t) V \ldots \W(t) V }$.
Each POVM involves only two observables, 
$\W(t)$ and $V$, not three observables.

Second, suppose that (i) $A, \ldots, R$ are unitary,
as well as Hermitian, and (ii) $\rho$ is pure.
$| \OTOCK |$ equals an overlap
$| \braket{ \psi'_{\rm II} }{ \psi'_{\rm I} } |$,
as $F(t)$ was shown to in the introduction.
Implementing 
$A(t_1)$, then $B(t_2)$, etc., then $E ( t_{ \BarK } )$
prepares $\ket{ \psi'_{\rm II} }$.
Implementing an analogous sequence 
prepares $\ket{ \psi'_{\rm II} }$.
The overlap $| \OTOCK |$ compares one sequence to the other,
rather than comparing all the observables 
that define the sequences.
An entropic uncertainty relation, in contrast,
reflects all the observables' disagreements with each other.}
Hence our entropic uncertainty relations
extend to arbitrary-point OTOCs.

%
%
\section{Entropic uncertainty relations for weak values beyond scrambling}
\label{sec:Beyond}

\emph{Weak values}, like OTOCs, involve time reversals
and measurement sequences~\cite{Aharonov_88_How,Dressel_14_Understanding}.
Consider preparing a quantum system 
in a state $\ket{i}$ at a time $t = 0$,
evolving the system for a time $t''$ under a unitary $U_{t''}$,
measuring a nondegenerate observable 
$F = \sum_f  f  \,  \ketbra{f}{f}$,
and obtaining the outcome $f$.
Let $A  =  \sum_a  a  \ketbra{a}{a}$ denote
a nondegenerate observable that fails to commute with $F$.

Which value can most reasonably be attributed, retrodictively,
to the $A$ at a time $t' \in (0, t'')$,
given that $\ket{i}$ was prepared and that
the measurement yielded $f$?
The \emph{weak value}
\begin{align}
   \label{eq:Weak_Val_Def}
   A_\weak  :=  \frac{ \bra{ f' }  A  \ket{ i' } }{ 
                        \braket{ f' }{ i' } }  \, ,
\end{align}
is the expectation value conditioned on 
the preselection and postselection.
$\ket{ f' }  :=  U_{t'' - t' }  \ket{f}$ and
$\ket{ i' }  :=  U_{ t' }  \ket{ i }$
denote time-evolved states.

Consider eigendecomposing $A$,
then factoring out the sum and eigenvalues.
Multiplying the numerator and denominator by $\braket{ i' }{ f' }$ yields
\begin{align}
   \label{eq:Aw_ito_KD}
   A_\weak (i, f)  
   =  \sum_a  a  \:  \frac{
   \braket{f'}{a}  \braket{a}{i'}  \braket{i'}{f'} }{ 
   p ( f | i ) }  \, ,
\end{align}
wherein $p ( f | i )  =  | \braket{ f' }{ i' } |^2$
denotes a conditioned probability.
The numerator is a \emph{Kirkwood-Dirac quasiprobability}~\cite{Kirkwood_33_Quantum,Dirac_45_On},
an extension of which is the OTOC quasiprobability~\cite{NYH_18_Quasi}.
The Kirkwood-Dirac quasiprobability governs
the conditional quasiprobability 
$\braket{f'}{a}  \braket{a}{i'}  \braket{i'}{f'}  /   p ( f | i )$
that, if $\ket{i}$ is prepared and 
the $F$ measurement yields $f$,
$a$ is the value most reasonably attributable to $A$ retrodictively.

$A_\weak$ generalizes to arbitrary initial states $\rho$
and to degenerate observables
$A  =  \sum_a  a  \,  \Pi^A_a$ and 
$F  =  \sum_f  f  \,  \Pi^F_f$:
\begin{align}
   \label{eq:Qw_Def_Gen}
   A_\weak  (\rho, f)
   & =   \frac{ \Tr  \left(  \Pi^{ F ( t'' - t' ) }_f  A  \rho( t' )  \right) }{   
   p ( f | \rho ) } \, .
\end{align}
The time-evolved state
$\rho (t' )  :=  U_{t'}^\dag  \rho  U_{t'}$,
and the conditional probability
$p ( f | \rho )  :=  \Tr  \left(  \Pi^{ F ( t'' - t' ) }_f  \rho(t')  \right)$.
One can infer $A_\weak$ experimentally
by preparing $\rho$, evolving the system for a time $t'$,
measuring $A$ weakly,
evolving the system for a time $t'' - t'$,
and measuring $F$ strongly.
One performs this protocol in many trials.
$A_\weak$ is inferred from the measurement statistics.

$A_\weak$ can range outside the spectrum of $A$,
as advertised in the foundational paper~\cite{Aharonov_88_How}.
Hence the physical significances of $A_\weak$ have galvanized debate (e.g.,~\cite{Ferrie_14_How,Vaidman_14_Comment,Cohen_14_Comment,Aharonov_14,Sokolovski_14_Comment,Brodutch_15_Comment,Ferrie_15_Ferrie,Dressel_15_Weak}).
Weak values have been interpreted in terms of 
conditioned expectation values~\cite{Aharonov_88_How}
and disturbances by measurements~\cite{Dressel_12_Significance}.
Kirkwood-Dirac quasiprobabilities have been interpreted in terms of 
operator decompositions~\cite{Lundeen_11_Direct,Lundeen_12_Procedure}
and Bayesian retrodiction~\cite{Aharonov_88_How,Johansen_04_Nonclassical,Hall_01_Exact,Hall_04_Prior,Dressel_15_Weak}.

We introduce another physical significance:
Weak values govern first-order-in-$g$ terms in entropic uncertainty bounds
for POVMs that involve weak measurements.
Kirkwood-Dirac quasiprobabilities play an analogous role 
in analogous bounds.
We present the results in Sec.~\ref{sec:Result_Unc_Rel_Aw_KD},
then illustrate with a qubit in Sec.~\ref{sec:Aw_Qubit_Ex}.


%
%
%
\subsection{Entropic uncertainty relations for weak values
and Kirkwood-Dirac quasiprobabilities}
\label{sec:Result_Unc_Rel_Aw_KD}

Consider a quantum system associated with
a Hilbert space $\Hil$.
Let $\rho  \in  \mathcal{D} ( \Hil )$ denote 
any state of the system.
Let $A = \sum_a  a  \,  \Pi^A_a$,  
$F = \sum_f  f  \,  \Pi^F_f$,
and $\I  =  \sum_i  \lambda_i  \,  \Pi^I_i$
be eigenvalue decompositions of observables.
(The index $i$ signifies ``initial'' and should not be confused with $\sqrt{ -1 }$.)

The uncertainty relation for $A_\weak$
features a POVM that we label I.
One measures $A$ weakly, then $F$ strongly:
$\Set{  \KrausI{j}{f}  :=  \Pi^F_f  \KrausGenBB{A}{j}  }$. 
The weak-measurement Kraus operator
$\KrausGenBB{A}{j}  =  \sqrt{ \pGen{A}{j} }  \, \id  +  \gGen{A}{j}  A
+ O \left( g^2 \right)$.
The $O \left( g^2 \right)$ signifies terms of second order in
the Hamiltonian's coupling parameter
(e.g., the $\tilde{g}$ in the spin-chain example
of Sec.~\ref{sec:Numerics}).
We define as POVM II a strong measurement of $\I$:
$\Set{ \KrausII{i}  :=  \Pi^\I_i  }$.

Define the entropies $H_\alpha \left(  A F  \right)_\rho$, and 
$H_\alpha ( \I )_\rho$
via analogy with the QI-scrambling entropies
(Sec.~\ref{sec:Entropies}).
One can infer the weak value\footnote{
We have tweaked our notation for $A_\weak$.
The first argument $i$, labels the subspace
over which the state $\Pi^\I_i  /  \Tr \left(  \Pi^I_i  \right)$
is maximally mixed.}
\begin{align}
   \label{eq:Aw_In_Thm}
   A_\weak(i, f)  
   =  \frac{  \Tr  \left(  \Pi^F_f  A  \Pi^\I_i  \right)  }{
                 \Tr  \left(  \Pi^F_f  \Pi^\I_i  \right)  \,  
                 \Tr  \left(  \Pi^\I_i  \right)  }
\end{align}
by preparing the state 
$\Pi^\I_i  /   \Tr  \left(  \Pi^\I_i  \right)$,
measuring $A$ weakly, and 
postselecting a strong $F$ measurement on $f$.

\begin{theorem}
\label{theorem:OTOC_free}
POVMs I and II obey entropic uncertainty relations
dependent on the weak value $A_\weak (i, f)$:
\begin{align}
   \label{eq:Weak_Unc_Rel}
   & H_\vN  (  \I )_\rho
   +  H_\vN \left(  A F  \right)_\rho  
   \geq  f_\weak
   \, ,  \quad \text{and}  \\
   \label{eq:Weak_Unc_Rel_Rastegin}
   & H_\alpha ( \I )_\rho
   +  H_\beta ( A F )_\rho  
   \geq  f_\weak \, .
\end{align}
The bound has the form
\begin{align}
   \label{eq:Bound_Beyond}
   f_\weak  
   :=   & \min_{i, j, f}  \Big\{ 
   - \log \left(  \pGen{A}{j}  \,   \Tr  \left(  \Pi^F_f  \Pi^\I_i  \right)  \right)
   \\ \nonumber & 
   -  \frac{2}{ \ln 2  }  \:
   \frac{ \Tr  \left(  \Pi^\I_i  \right) }{ \sqrt{ \pGen{A}{j} } }  \,
   \Real  \left(  \gGen{A}{j}    A_\weak(i, f)   \right)
   + O \left( g^2 \right)  \Big\} \, .
\end{align}
The R\'enyi orders $\alpha$ and $\beta$ satisfy
$\frac{1}{\alpha}  +  \frac{1}{\beta}  =  2$, and
$\rho$ denotes an arbitrary state.
\end{theorem}

The proof is analogous to the proof of Theorem~\ref{theorem:OTOC_thm_Toma}.
The forward and reverse POVMs 
are replaced with POVMs I and II.
One can prove analogous uncertainty relations in which
Kirkwood-Dirac quasiprobabilities replace $A_\weak$.
The weak measurement of $A$ gives way to
a weak measurement of an $A$ eigenprojector.
The uncertainty bound~\eqref{eq:Weak_Unc_Rel_Rastegin}
can be smoothed when $(\alpha, \beta) = (\infty, 1/2)$.
For uncertainty relations that involve weak measurements,
but are not entropic, see~\cite{Hall_16_Products}.

%
%
\subsection{Qubit example}
\label{sec:Aw_Qubit_Ex}

Let us illustrate the uncertainty relation~\eqref{eq:Weak_Unc_Rel_Rastegin} 
for $(\alpha, \beta) = (\infty, 1/2)$.
The system, denoted by a subscript $\sys$,
consists of a qubit.
So does the detector, denoted by $\detect$.
Let $\I  =  \sigma^z_\sys$, $A = \sigma^y_\sys$, and 
$F = \sigma^x_\sys$.

The weak measurement manifests as follows:
The detector begins in the state $\ket{ x+ }$,
a $z$-controlled $y$ couples the system
to the detector weakly,
and the detector's $\sigma^y_\detect$ is measured strongly.
The weak values $A_\weak (z_\sys, x_\sys)
=  x_\sys  z_\sys  i$
are imaginary and so nonclassical~\cite{Dressel_12_Significance}:
$A$ has only real eigenvalues $a$,
but the conditioned average $A_\weak$ is imaginary.

We illustrate the uncertainty relation's LHS with
$\rho  =  \ketbra{ z+ }{ z+ }$.
The inequality is calculated in App.~\ref{app:Qubit_Aw}:
$2.00  \geq  2.00  -  \frac{2}{ \ln 2 }  \,  | \tilde{g} |  
+  O  \left(  \tilde{g}^2  \right)$.
If $\tilde{g}  =  2.00 \times 10^{ -2 }$, as in Sec.~\ref{sec:Numerics}, 
the relation approximates to  $2.00  \geq  1.94$.
The bound is satisfied and is tight at order $g^0$.

%
%
\section{Discussion}
\label{sec:Discussion}

We have reconciled two measures of disagreement
between quantum operators:
entropic uncertainty relations and out-of-time-ordered correlators (OTOCs).
The reconciliation unites several subfields of physics: 
(i) quasiprobabilities and weak measurements 
tie (ii) quantum information theory 
to (iii) condensed matter and (iv) high-energy physics.
Information theory and complexity theory 
have begun intersecting with 
condensed matter and high-energy physics recently,
shedding light on black holes, information propagation,
and space-time (e.g.,~\cite{Hayden_07_Black,Swingle_12_Entanglement,Harlow_13_Quantum,Pastawski_15_Holographic,Bao_15_Holographic,Hosur_16_Chaos,Brown_18_Second}).
This paper broadens the intersection 
into quasiprobability and quantum-measurement theory 
and farther into quantum information theory.

This broadening has two more important significances:
one for OTOC theory and one for weak-measurement theory.
First, the extension reconciles the OTOC's $V$ with
the tiny perturbation that triggers violent consequences
in the classical butterfly effect:
$V$ can naturally be regarded, our uncertainty relations show,
as being measured weakly.
The weak measurement is perturbative literally,
in the coupling strength $g$.

Within measurement theory, second,
we have uncovered a physical significance 
of weak values $A_\weak$ and 
Kirkwood-Dirac quasiprobabilities:
These quantities govern first-order terms in
entropic uncertainty relations obeyed by weak measurements.
Quantum information theory therefore sheds light on
mathematical objects whose interpretations have been debated in
quantum optics, quantum foundations, and quantum computation.

In a recent paper, an uncertainty relation was extended to unitaries,
then applied to bound the OTOC~\cite{Bong_18_Strong}.
OTOC bounds have been known to limit the speed at which 
many-body entanglement can develop~\cite{Maldacena_15_Bound,Sekino_08_Fast,Lashkari_13_Towards}.
The present work takes a fundamentally different approach:
Scrambling takes central stage in this paper,
whose main purpose is to unite 
two communities' notions of quantum operator disagreement.
Additionally, our uncertainty relations are entropic,
tapping into recent developments in pure quantum information theory.
Finally, our formalism covers both unitary and Hermitian
OTOC operators $V$ and $\W$.

This work uncovers several research opportunities.
Inspired by condensed matter, we have focused on discrete systems.
Also continuous systems---quantum field theories (QFTs)---have OTOCs
used to study, e.g., black holes in the anti-de-Sitter-space/conformal-field-theory (AdS/CFT) duality~\cite{Kitaev_15_Simple,Shenker_Stanford_14_BHs_and_butterfly,Shenker_Stanford_14_Multiple_shocks,Roberts_15_Localized_shocks,Roberts_Stanford_15_Diagnosing,Maldacena_15_Bound,Haehl_16_Schwinger_I,Roberts_16_Lieb}.
Entropic uncertainty relations for continuous-variable systems 
have been derived~\cite{Bialynicki_Birula_75_Uncertainty,Beckner_75_Inequalities,Babenko_61_Inequality,Hirschman_57_Note,Guanlei_09_Generalized,Huang_11_Entropic,Bialynicki_Birula_11_Entropic,Bialynicki_Birula_06_Formulation}.
They should be applied to characterize scrambling in QFTs.

Second, Theorems~\ref{theorem:OTOC_thm_Toma} and~\ref{theorem:OTOC_free}
can be tested experimentally.
The techniques needed exist:
OTOC measurements have been proposed in detail~\cite{Swingle_16_Measuring,Yao_16_Interferometric,Zhu_16_Measurement,Bohrdt_16_Scrambling,Campisi_16_Thermodynamics,Tsuji_17_Exact,NYH_17_Jarzynski,NYH_18_Quasi,Tsuji_18_Out},
and early-stage OTOC-measurement experiments have performed~\cite{Li_16_Measuring,Garttner_16_Measuring,Wei_16_Nuclear,Meier_17_Exploring};
weak values and Kirkwood-Dirac distributions
have been measured weakly~\cite{Ritchie_91_Realization,Pryde_05_Measurement,Bollen_10_Direct,Lundeen_11_Direct,Lundeen_12_Procedure,Groen_13_Partial,Bamber_14_Observing,Mirhosseini_14_Compressive,Smith_04_Continuous,Hacohen_16_Quantum,White_16_Preserving,Chen_18_Experimental};
and entropic uncertainty relations have been tested experimentally~\cite{Sulyok_15_Experimenetal,Berta_16_Entropic,Xing_17_Experimental,Xiao_17_Quantum}.
Testing Theorem~\ref{theorem:OTOC_thm_Toma}
should be feasible in the immediate future,
especially through the weak-measurement proposal 
for inferring the OTOC quasiprobability $\SumKD{\rho}$~\cite{NYH_17_Jarzynski,NYH_18_Quasi}.
Prospective platforms include superconducting qubits, ultracold atoms, 
trapped ions, quantum dots, and potentially NMR.

Testing Theorem~\ref{theorem:OTOC_free} experimentally
requires even fewer resources:
Interacting many-body systems are unnecessary,
and one weak measurement per trial suffices.
Tantalizingly, though, two~\cite{Piacentini_16_Measuring,Suzuki_16_Observation,Thekkadath_16_Direct}
and three~\cite{Chen_18_Experimental} 
sequential weak measurements have been realized recently.
They can be applied to 
(i) characterize higher-order terms in 
Eqs.~\eqref{eq:Weak_Unc_Rel_Rastegin}
and~\eqref{eq:Bound_Beyond},
(ii) test entropic uncertainty relations 
for higher-point OTOCs (Sec.~\ref{sec:K_OTOCs}),
and (iii) test entropic uncertainty relations for
POVMs of sequential weak measurements.

Third, the entropic uncertainty relations for scrambling
can be smoothed with an error tolerance $\varepsilon$.
When smoothing, one ignores highly unlikely events~\cite{Renner_05_Security}.
Highly unlikely outcomes of weak-measurement experiments
correspond to anomalous weak values
and nonclassical quasiprobability values~\cite{Dressel_15_Weak}.
Nonclassical operator disagreement
underlies nontrivial uncertainty relations.
Whether smoothing trivializes 
entropic uncertainty relations for weak measurements merits study.
Rough numerical studies suggest that 
$\varepsilon$ might actually tighten 
the spin-chain bound~\eqref{eq:Rastegin_OTOC}.
%

Like smoothing, conditioning generalizes
the entropic uncertainty relations in~\cite{Tomamichel_12_Framework}.
Consider holding a memory $\sigma$ that is entangled with
a to-be-measured state $\rho$.
Conditioning on $\sigma$ can change
your uncertainty about the measurement outcome.
Certain scrambling setups might be cast in terms of a memory $\sigma$.
An example consists of a qubit chain and an ancilla qubit~\cite{Swingle_Notes}.
Consider entangling the ancilla with the chain's central qubit,
then evolving the chain under a many-body Hamiltonian.
The entanglement with the ancilla spreads through the chain.
The ancilla might be cast as the memory $\sigma$
in conditioned entropic uncertainty relations for scrambling.

Finally, nonclassicality of 
$\SumKD{\id}$ and $A_\weak$
might strengthen the uncertainty bounds.
The quasiprobability behaves nonclassically
by acquiring negative real and nonzero imaginary components.
The weak value $A_\weak$ behaves nonclassically
by lying outside the spectrum of $A$.
Such nonclassical mathematical behavior
can signal nonclassical physics~\cite{Gross_05_Computational,Galvao_05_Discrete,Spekkens_08_Negativity,Veitch_12_Negative,Howard_14_Contextuality,Delfosse_15_Wigner}.
The quasiprobability's nonclassicality features little 
in our numerical example (Sec.~\ref{sec:Numerics}):
First, the quasiprobability's imaginary part vanishes
when evaluated on $\id$~\cite[Sec.~III and Sec.~V A]{NYH_18_Quasi}.
Hence $\Imag  \left(  \SumKD{\id}  \right)$ 
cannot influence the bound.
Second, $\SumKD{\id}$ assumes negative values,
but not when $w_1 = w_2$.
Higher-point-OTOC quasiprobabilities 
could avoid this roadblock,
and assume negative values in the bound,
as higher-point forward and reverse protocols 
depend on weak $\W(t)$ measurements
(Sec.~\ref{sec:K_OTOCs}). 
Nonclassicality's potential to tighten uncertainty bounds
merits study.

%
%
%



%
%
\vspace{1em}
\begin{acknowledgments}
We are grateful for conversations with Fernando G. S. L. Brand\~{a}o, Sean Carroll, Justin Dressel, Patrick Hayden, Jos\'e Ra\'ul Gonzalez Alonso, Renato Renner, Brian Swingle, and Marco Tomamichel.
NYH is grateful for support from the Institute for Quantum Information and Matter (IQIM),
for a Barbara Groce Graduate Fellowship,
and for a Graduate Fellowship from the Kavli Institute for Theoretical Physics.
NYH acknowledges Mark van Raamsdonk, his fellow conference organizers, the ``It from Qubit'' collaboration, and UBC for their hospitality and their invitation to participate in ``Quantum Information in Quantum Gravity III,'' where this project partially took shape.
AB acknowledges support from the Walter Burke Institute for Theoretical Physics and the U.S. Department of Energy, Office of Science, Office of High Energy Physics, under Award Number DE-SC0011632.
JP is supported partially by the Simons Foundation and 
partially by the Natural Sciences and Engineering Research Council of Canada.
The IQIM is an NSF Physics Frontiers Center (NSF Grant PHY-1125565) 
that receives support from 
the Gordon and Betty Moore Foundation (GBMF-2644).
The KITP is supported by the NSF under Grant No. NSF PHY-1125915.
\end{acknowledgments}

\begin{appendices}

\onecolumngrid

\renewcommand{\thesection}{\Alph{section}}
\renewcommand{\thesubsection}{\Alph{section} \arabic{subsection}}
\renewcommand{\thesubsubsection}{\Alph{section} \arabic{subsection} \roman{subsubsection}}

\makeatletter\@addtoreset{equation}{section}
\def\theequation{\thesection\arabic{equation}}

\section{Proof of Theorem~\ref{theorem:OTOC_thm_Toma}}
\label{section:Proof_MainThm}

Tomamichel presents an entropic uncertainty relation 
for the smooth entropies
$H_\Min^\varepsilon$ and $H_\Max^\varepsilon$~\cite[Result~7]{Tomamichel_12_Framework};
Krishna and Parthasarathy derive one for 
$H_\vN$ and $H_\vN$~\cite[Corollary~2.6]{Krishna_01_Entropic};
and Rastegin presents one for
$H_\alpha$ and $H_\beta$~\cite[Ineq.~(13)]{Rastegin_08_Uncertainty}
(proved in~\cite{Rastegin_08_Statement}).
We use Tomamichel's notation, for concreteness.
But the three uncertainty relations have the same RHSs.
Hence our use of~\cite[Result~7]{Tomamichel_12_Framework}
translates directly into uses of the other two bounds.

Tomamichel considers POVMs $\mathcal{X}$ and $\mathcal{Y}$
whose outcomes are recorded in classical registers $X$ and $Y$.
The systems $B$ and $C$ can hold quantum side information about,
or have correlations with, $X$ and $Y$.
An agent performing an information-processing task
might wish to infer about $X$ and $Y$,
given access to $B$ and $C$.
Tomamichel presents the entropic uncertainty relation
\begin{align}
   \label{eq:TomaTheorem}
   H_\Min^\varepsilon ( X | B )_\rho  
   +  H_\Max^\varepsilon ( Y | C )_\rho
   \geq  \log  \frac{1}{ c ( \mathcal{X} ,  \mathcal{Y} ) }  
\end{align}
in~\cite{Tomamichel_12_Framework} (see also~\cite{Tomamichel_12_Tight,Thinh_12_Tomographic,Berta_16_Smooth}).
The smooth entropies $H_\Min^\varepsilon$ and $H_\Max^\varepsilon$
follow from extremizing $H_\Min$ and $H_\Max$.
The POVM overlap $c$, defined in Eq.~\eqref{eq:Overlap1},
generalizes the overlap~\eqref{eq:Overlap_Def}.

We set the smoothing parameter $\varepsilon$ to zero.
We also trivialize the conditioning,
setting the states of $B, C  \propto  \id$.
Let us substitute in the POVMs~\eqref{eq:Fwd_POVM} and~\eqref{eq:Rev_POVM}:
\begin{align}
   \label{eq:Proof_Help1} &  
   H_\Min  \LParen  V  \W(t)  \RParen_\rho
   +  H_\Max  \LParen  \W(t)  V  \RParen_\rho
   \geq  -  \log  \LParen  c  \left(  
   \Set{  \POVMF{ v_1 }{ j_1 }{ w_1 }  } ,
   \Set{  \POVMR{ v_2 }{ j_2 }{ w_2 }  }
   \right)  \RParen  \, .
\end{align}
The POVM overlap $c$ is defined as\footnote{
Reference~\cite{PhysRevA.89.022112} strengthens 
the $H_\vN$--$H_\vN$ bound 
by replacing $c ( \mathcal{X} ,  \mathcal{Y} )$ with 
$c^\prime ( \mathcal{X} ,  \mathcal{Y} )
:= \Min\Set{\Max_x \|  
\sum_y \mathcal{Y}^y
\mathcal{X}^x\mathcal{Y}^y \|,
\Max_y \|
\sum_x \mathcal{X}^x \mathcal{Y}^y \mathcal{X}^x\|}$.
(See Sec.~III.D of~\cite{Coles_17_Entropic} for a review.) 
That is, $c$ may be replaced with 
$c^\prime\le c$ in 
the $H_\vN$--$H_\vN$ version of (\ref{eq:TomaTheorem}). 
(We thank an anonymous reviewer for bringing this result to our attention.) 
Substituting in our POVMs and Taylor-approximating, 
as below, would be straightforward. 
However, the resulting bound would involve 
more-complicated operators
than our uncertainty bound for scrambling.
Identifying OTOC quasiprobabilities in the strengthened bound 
may therefore be more difficult.
Re-engineering the POVMs might enable one to strengthen the bound
while retaining the bound's dependence on the OTOC quasiprobability
and so the bound's tightening at the scrambling time.}
\begin{align}
   \label{eq:Overlap1}
   c  \left(  \Set{  \POVMF{ v_1 }{ j_1 }{ w_1 }  }  ,
                \Set{  \POVMR{ v_2 }{ j_2 }{ w_2 }  }  \right)
   & :=  \max_{ j_1 , j_2,  w_1 , w_2 }  \Set{
   \left\|  \KrausF{ v_1 }{ j_1 }{ w_1 }
            \KrausR{ v_2 }{ j_2 }{ w_2 }  \right\|^2  }  \, .
\end{align}
The operator norm has the form
\begin{align}
   \label{eq:Op_Norm1}
   & \left\|  \KrausF{ v_1 }{ j_1 }{ w_1 }
            \KrausR{ v_2 }{ j_2 }{ w_2 }  \right\|
   =  \lim_{ \alpha \to \infty }  \left\{
   \Tr  \left(  \sqrt{  
   \left[  \KrausF{ v_1 }{ j_1 }{ w_1 }  \KrausR{ v_2 }{ j_2 }{ w_2 }  \right]^\dag
   \left[  \KrausF{ v_1 }{ j_1 }{ w_1 }  \KrausR{ v_2 }{ j_2 }{ w_2 }  \right]
   }^{ ^{ \; \alpha } }  \right)  \right\}^{ 1 / \alpha }  \, .
\end{align}
The outer square-root equals, by
Eqs.~\eqref{eq:Fwd_POVM} and~\eqref{eq:Rev_POVM},
\begin{align}
   \sqrt{   \KrausR{ v_2 }{ j_2 }{ w_2 }^\dag  
   \KrausF{ v_1 }{ j_1 }{ w_1 }^\dag  
   \KrausF{ v_1 }{ j_1 }{ w_1 }  \KrausR{ v_2 }{ j_2 }{ w_2 }  } 
   & =  \sqrt{  \ProjWt{w_2}     \KrausV{v_2}{j_2}  
   \left(  \KrausV{v_1}{j_1}  \right)^\dag
   \ProjWt{w_1}
   \KrausV{v_1}{j_1}
   \left(  \KrausV{v_2}{j_2}  \right)^\dag
   \ProjWt{w_2}  }
   \\ &
   \equiv  \sqrt{ O }  \, .
\end{align}
The two central projectors have collapsed into one:
$\left(  \ProjWt{w_1}  \right)^2  =  \ProjWt{w_1}$.

The operator $O$ is Hermitian and so eigendecomposes.
The eigenvalues are real and nonnegative, being
the squares of the singular values of 
$\KrausF{ v_1 }{ j_1 }{ w_1 }  \KrausR{ v_2 }{ j_2 }{ w_2 }$.
Also a physical argument implies
the eigenvalues' reality and nonnegativity:
$O$ is proportional to a quantum state:
$\ProjWt{w_2} / \Tr  \left(  \ProjWt{w_2}  \right)$
represents the state that is maximally mixed over
the eigenvalue-$w_2$ eigenspace of $\W(t)$.
Imagine preparing $\ProjWt{w_2} / \Tr  \left(  \ProjWt{w_2}  \right)$,
subjecting the state to the quantum channel defined by
the operation elements~\cite{NielsenC10}
$\Set{  \left(  \KrausV{v_2}{j_2}  \right)^\dag  }_{j_2}$,\footnote{
We must prove that 
$\Set{  \left(  \KrausV{v_2}{j_2}  \right)^\dag  }_{j_2}$
defines a quantum channel.
$\Set{  \KrausV{v_2}{j_2}  }_{j_2}$ does by definition, so
each $\KrausV{v_2}{j_2}$ maps 
the input Hilbert space to the output Hilbert space, and 
$\sum_{j_2}   \left(  \KrausV{v_2}{j_2}  \right)^\dag    \KrausV{v_2}{j_2}
=  \id$.
The operator $\KrausV{v_2}{j_2}$ differs from
$\left(  \KrausV{v_2}{j_2}  \right)^\dag$ only by
complex conjugation of the coupling $\gV{j_2}  \in  \mathbb{C}$.
Hence $\sum_{j_2}  \KrausV{v_2}{j_2}   
\left(  \KrausV{v_2}{j_2}  \right)^\dag  =  \id$,
as required of Kraus operators.
This mathematical result complements physical intuition:
Suppose that the detector manifests as a qubit.
A common interaction rotates the detector's state
conditionally on the system's state~\cite{Dressel_14_Implementing,NYH_18_Quasi,Dressel_18_Strengthening}.
Let $\Set{  \KrausV{v_2}{j_2}  }_{j_2}$
follow from a rotation in some fiducial direction.
$\Set{  \left(  \KrausV{v_2}{j_2}  \right)^\dag  }_{j_2}$
follows from a rotation in the opposite direction.
Now, suppose that the detector manifests as 
a particle in some potential.
A common interaction conditionally kicks the detector.
If $\Set{  \KrausV{v_2}{j_2}  }_{j_2}$
follows from a kick in one direction,
$\Set{  \left(  \KrausV{v_2}{j_2}  \right)^\dag  }_{j_2}$
follows from a kick in the opposite.}
subjecting the state to the channel defined by
$\Set{  \KrausV{v_1}{j_1}   }_{j_1}$, 
and then measuring $\W(t)$ projectively.
The resultant state, $\sigma_f$, is proportional to $O$.
The proportionality factor equals $\Tr (O)$,
the joint probability that 
(i) this realization of the initial channel's action
is labeled by $j_2$,
(ii) this realization of the second channel's action
is labeled by $j_1$, and
(iii) the $\W(t)$ measurement yields outcome $w_2$.
Since $\sigma_f  =  O / \Tr (O)$ 
$\sigma_f$ is positive semidefinite
and $\Tr (O)$ equals a probability,
the eigenvalues of $O$ are real and nonnegative.

The eigenvectors of $O$ are
eigenvectors of $\ProjWt{w_2}$.
$\ProjWt{w_2}$ has two distinct eigenvalues $\eta$:
$\eta = 0$, of degeneracy
$\Tr  \left(  \id  -  \ProjWt{w_2}  \right)$,
and $\eta = 1$, of degeneracy
$\Tr  \left(  \ProjWt{w_2}  \right)$.
Let $\Lambda_\eta^r$ denote
the $r^\th$ $O$ eigenvalue 
associated with any eigenvector
in the $\eta$ eigenspace of $\ProjWt{w_2}$.
If $d_\eta$ denotes the degeneracy of $\Lambda_\eta^r$,
$r  =  1 , 2, \ldots  d_\eta$.
(We have omitted the $\eta$ dependence 
from the symbol $r$ for notational simplicity.)
Every eigenvalue-0 eigenvector of $\ProjWt{w_2}$ is
an eigenvalue-0 eigenvector of $O$:
$\Lambda_0^r  =  0  \;  \;  \forall 
r  =  1, 2, \ldots  \Tr \left(  \id  -  \ProjWt{w_2}  \right)$.
Hence $O$ eigendecomposes as
\begin{align}
   \label{eq:O_Help1}
   O  & =  \sum_{ \eta = 0}^1  \sum_{r = 1}^{ d_\eta }
   \Lambda_\eta^r  \,  \Pi_\eta^r  
   =  0  \left(  \id  -  \ProjWt{w_2}  \right)
   +  \sum_{ r = 1 }^{ d_1 }  \Lambda_1^r  \,  \Pi_1^r  \, .
\end{align}

We use this eigenvalue decomposition
to evaluate the RHS of Eq.~\eqref{eq:Op_Norm1},
working from inside to outside.
The outer square-root has the form
$\sqrt{ O }  
   =  \sum_{r = 1}^{ d_1 }  \sqrt{ \Lambda_1^r }  \,  \Pi_\eta^r  \, .$
The projectors project onto orthogonal subspaces, so
$\left(  \sqrt{O}  \right)^\alpha
   =  \sum_{ r = 1}^{ d_1 }
   \left(  \Lambda_1^r  \right)^{ \alpha / 2 }  \,
   \Pi_1^r  \, .$
We take the trace,
$\Tr  \left(  \left[  \sqrt{O}  \right]^\alpha  \right)
=  \sum_{r = 1}^{ d_1 }
\left(  \Lambda_1^r  \right)^{ \alpha / 2 }$,
then exponentiate:
$\left\{  \Tr  \left(  \left[  
\sqrt{O}  \right]^\alpha  \right)  \right\}^{ 1 / \alpha }
=  \left[  \sum_{r = 1}^{ d_1 }
\left(  \Lambda_1^r  \right)^{ \alpha / 2 }
\right]^{ 1 / \alpha }$.
The limit as $\alpha \to \infty$ gives 
the RHS of Eq.~\eqref{eq:Op_Norm1}:
\begin{align}
   \label{eq:O_Help2}
   \left\|  \KrausF{ v_1 }{ j_1 }{ w_1 }
            \KrausR{ v_2 }{ j_2 }{ w_2 }  \right\|
   & =  \lim_{ \alpha \to \infty }  \left\{  \Tr  
   \left(  \left[  \sqrt{O}  
   \right]^\alpha  \right)  \right\}^{ 1 / \alpha }  \\
    \label{eq:O_Help3}
   & =  \lim_{ \alpha \to \infty }
   \left[  \sum_{r = 1}^{ d_1 }
   \left(  \Lambda_1^r  \right)^{ \alpha / 2 }
   \right]^{ 1 / \alpha }  \, .
\end{align}

Only the greatest eigenvalue to survives:
$\left\|  \KrausF{ v_1 }{ j_1 }{ w_1 }
            \KrausR{ v_2 }{ j_2 }{ w_2 }  \right\|
  =  \sqrt{ \Lambda_1^\Max }$.
But $\Lambda_1^\Max$ is neither 
a parameter chosen by the experimentalist
nor obviously experimentally measurable.
Hence bounding the entropies with $\Lambda_1^\Max$
is useless.

Probabilities and quasiprobabilities are measurable.
$\Tr (O)$ equals a combination of probabilities and quasiprobabilities.
We therefore seek to shift the $\Tr$ of Eq.~\eqref{eq:O_Help2}
inside the $[ . ]^\alpha$ and the $\sqrt{ . }$ \, .
Equivalently, we seek to shift 
the $\sum$ of Eq.~\eqref{eq:O_Help3}
inside the $( . )^{ \alpha / 2 }$.
We do so at the cost of introducing an inequality:
\begin{align}
   \label{eq:RevTri}
   \sum_r  ( \Lambda_1^r )^{ \alpha / 2 } 
   \leq  \left(  \sum_r  \Lambda_1^r  \right)^{ \alpha / 2 }  
\end{align}
for all $\alpha /2  \geq 1$.
This inequality follows from 
the Schatten $p$-norm's monotonicity.
The Schatten $p$-norm of an operator $\sigma$ is defined as
$|| \sigma ||_p  :=  \left[  \Tr  \left(  \sqrt{ 
\sigma^\dag  \sigma  }^{ \, p }  \right)  \right]^{ 1 / p }$,
for $p \in [1, \infty)$.
As $p$ increases, the Schatten norm decreases monotonically:
\begin{align}
   \label{eq:Schatten_Mon}
   || \sigma ||_p  \leq  || \sigma ||_q  
   \quad \text{if}  \quad   p  \geq  q  \, .
\end{align}
Let $p = \alpha / 2$ and $q = 1$.
Raising each side of Ineq.~\eqref{eq:Schatten_Mon}
to the $\alpha / 2$ power
yields Ineq.~\eqref{eq:RevTri}.
Applying Ineq.~\eqref{eq:RevTri} to Eq.~\eqref{eq:O_Help3}
bounds the operator norm as
\begin{align}
   \label{eq:Op_Norm2}
   \left\|  \KrausF{ v_1 }{ j_1 }{ w_1 }
            \KrausR{ v_2 }{ j_2 }{ w_2 }  \right\|
   & \leq  \sqrt{  \Tr  \left(  
   \ProjWt{w_2}  \KrausV{v_2}{j_2}  
   \left[  \KrausV{v_1}{j_1}  \right]^\dag
   \ProjWt{w_1}
   \KrausV{v_1}{j_1}
   \left[  \KrausV{v_2}{j_2}  \right]^\dag  \right)  } \, .
\end{align}
We have invoked the trace's cyclicality and
$\left(  \ProjWt{w_2}  \right)^2  =  \ProjWt{w_2}$.

Substituting into Eq.~\eqref{eq:Overlap1} bounds the overlap:
\begin{align}
   \label{eq:Overlap2}
   c  \left(  \Set{  \POVMF{ v_1 }{ j_1 }{ w_1 }  }  ,
                \Set{  \POVMR{ v_2 }{ j_2 }{ w_2 }  }  \right)
   & \leq  \max_{ j_1 , j_2,  w_1 , w_2 }
   \left\{  \Tr  \left(
   \ProjWt{w_2}  \KrausV{v_2}{j_2}  
   \left[  \KrausV{v_1}{j_1}  \right]^\dag
   \ProjWt{w_1}
   \KrausV{v_1}{j_1}
   \left[  \KrausV{v_2}{j_2}  \right]^\dag
   \right)  \right\}  \, .
\end{align}
We substitute into the trace from 
Eqs.~\eqref{eq:Fwd_POVM} and~\eqref{eq:Rev_POVM}:
\begin{align}
   \label{eq:Tr_In_C_1}
   & c  \left(  \Set{  \POVMF{ v_1 }{ j_1 }{ w_1 }  }  ,
                \Set{  \POVMR{ v_2 }{ j_2 }{ w_2 }  }  \right)
   \leq  \max_{ j_1 , j_2, w_1 , w_2 }
   \Big\{  \Tr  \Big(
   \ProjWt{w_2}  
   \left[  \sqrt{ \pV{j_2}  }  \,  \id  +  \gV{j_2}  \,  \ProjV{v_2}  \right]
   \\ \nonumber & \qquad \times
   \left\{  \sqrt{  \pV{j_1}  }  \,  \id  
             +  \left[  \gV{j_1}  \right]^*  \,  \ProjV{v_1}  \right\}
   \ProjWt{w_1}
   \left[  \sqrt{  \pV{j_1}  }  \,  \id  +  \gV{j_1}  \,  \ProjV{v_1}  \right]
   \left\{  \sqrt{  \pV{j_2}  }  \,  \id
             +  \left[  \gV{j_2}  \right]^*  \,  \ProjV{v_1}  \right\}
   \Big)  \Big\}  \, .
\end{align}
Multiplying out yields
\begin{align}
   \label{eq:Tr_In_C_2}
   & c  \left(  \Set{  \POVMF{ v_1 }{ j_1 }{ w_1 }  }  ,
                \Set{  \POVMR{ v_2 }{ j_2 }{ w_2 }  }  \right)
   \leq  \max_{ j_1 , j_2, w_1 , w_2 }   \Big\{  
   \pV{j_1}  \,  \pV{j_2}  \,  
   \Tr \left(  \ProjW{w_2}  \right)  \,  \delta_{ w_1  w_2  }
   \\ \nonumber & \quad
   +  \Big[  2 \sqrt{  \pV{j_1}  }  \:  \pV{j_2}
   \Real  \left(  \gV{ j_1 }  \right)
   \Tr  \left(  \ProjWt{ w_2 }  \ProjV{ v_1 }  \right)
   \delta_{ w_1 w_2 }
   +  2  \pV{ j_1 }  \sqrt{ \pV{ j_2 } }  \:
   \Real  \left(  \gV{ j_2 }  \right)
   \Tr  \left(  \ProjW{ w_2 }  \ProjV{ v_2 }  \right)
   \delta_{ w_1 w_2 }  \Big]
   \nonumber \\ & \quad
   +  \Big[  \pV{ j_2 }  \Verts{ \gV{ j_1 }  }^2
   \Tr  \left(  \ProjWt{ w_2 }  \ProjV{ v_1 }  \ProjWt{ w_1 }  \ProjV{ v_1 }  \right)
   +  \pV{ j_1 }  \Verts{ \gV{ j_2 }  }^2
   \Tr  \left(  \ProjWt{ w_2 }  \ProjV{ v_2 }  \ProjWt{ w_1 }  \ProjV{ v_2 }  \right)
   \nonumber \\ & \quad
   + 2  \sqrt{ \pV{ j_1 }  \pV{ j_2 }  }  
   \Real  \left(  \gV{ j_1 }  \gV{ j_2 }  
                   \Tr  \left(  \ProjWt{ w_2 }  \ProjV{ v_2 }  \ProjWt{ w_1 }  \ProjV{ v_1 }  \right)  
           \right)
   + 2 \sqrt{ \pV{ j_1 }  \pV{ j_2 }  }
   \Real \left(  \gV{ j_1 }  \left[ \gV{ j_2 } \right]^*  \right)
   \Tr  \left(  \ProjWt{ w_2 }  \ProjV{ v_1 }  \right)
   \delta_{ v_1 v_2 }  \delta_{ w_1 w_2 }
   \nonumber \\ & \quad
   + 2 \sqrt{ \pV{ j_2 }  }   \Verts{ \gV{ j_1 } }^2
   \Real \left(  \gV{ j_2 }  
   \Tr  \left(  \ProjWt{ w_2 }  \ProjV{ v_2}  \ProjWt{ w_1 }  \ProjV{ v_1 }  \right)  \right)
   \delta_{ v_1 v_2 }
   \nonumber \\ & \quad \nonumber
   + 2 \sqrt{ \pV{ j_1 } }  \Verts{ \gV{ j_2 } }^2
   \Real  \left(  \gV{ j_1 }
                    \Tr  \left(  \ProjWt{ w_2 }  \ProjV{ v_2 }  \ProjWt{ w_1 }  \ProjV{ v_1 }  \right)
          \right)   \delta_{ v_1 v_2 }  
   \Big]
   + \Verts{ \gV{ j_1 }  }^2      \Verts{  \gV{ j_2 }  }^2
   \Tr  \left(  \ProjWt{ w_2 }  \ProjV{ v_2 }  \ProjWt{ w_1 }  \ProjV{ v_1 }  \right)
   \delta_{ v_1 v_2 }
   \Big\}  \, .
\end{align}
Six of the traces are instances of $\SumKD{ \id }$.

Only the first term is constant in $g$.
If $g$ is small, therefore, 
the maximum obtains where the first term maximizes,
where $w_1 = w_2$.
Hence every RHS term 
is implicitly evaluated at $w_1 = w_2$.

We take the log of each side of Ineq.~\eqref{eq:Tr_In_C_2}.
The log's monotonicity implies
$\log c  \leq  \max \Set{ \log  (\ldots )  }$.
We negate each side, then shift the negative sign across the max
(as negative logs evoke entropies):
$- \log c  \geq  - \max \Set{  \log ( \ldots ) }
=  \min  \Set{ - \log ( \ldots ) }$.
With this inequality and with Ineq.~\eqref{eq:Tr_In_C_2},
we bound the RHS of Ineq.~\eqref{eq:TomaTheorem}.

Next, we factor out the $\pV{ j_1}  \,  \pV{ j_2 }  \,  
\Tr \left(  \ProjW{w_2}  \right)$
and invoke the log law for multiplication:
$ \min \Set{  
-  \log \left( \pV{ j_1}  \,  \pV{ j_2 }  \,   
\Tr \left(   \ProjW{w_2}   \right)  \right)
+ \log \left( 1 + [ \text{terms small in $g$} ]  \right)  }$.
We then Taylor-approximate in the $g$'s.
The quasiprobability values are assumed to be small enough
not to undermine the Taylor approximation.
This assumption is reasonable:
OTOC quasiprobability values $>1$ have not been observed 
in any of the numerical simulations performed
for this paper or for~\cite{NYH_18_Quasi}.
Moreover, the $g$'s can always be weakened enough
to offset any largeness of $\SumKD{\id}$.

%
%
%
\section{Analytical calculations for the spin-chain example}
\label{sec_Spin_Chain_Analytics}

Let us derive the results in Sec.~\ref{sec:Spin_Chain_Calcs}.
We calculate the detector probability 
$\pVH{j_\ell}  \equiv  \pVH{x_\ell}$,
the weak-measurement Kraus operators 
$\KrausVH{v_\ell}{j_\ell}  \equiv  \KrausVH{v_\ell}{x_\ell}$,
the coupling strengths 
$\gVH{j_\ell}  \equiv  \gVH{x_\ell}$,
and the entropies $H_\alpha$.

\textbf{Detector probability $\pVH{j_\ell}  \equiv  \pVH{x_\ell}$:}
Consider preparing the detector in $\ket{D}$,
then measuring $\hat{x}$.
The measurement has a probability
$\pVH{x_\ell}  L  =  |  \braket{ x_\ell }{ D }  |^2  L$
of yielding a position within $L$ of  $x_\ell$.
By Eq.~\eqref{eq:Det_State},
\begin{align}
   \label{eq:Ex_Prob_App}  
   \pVH{x_\ell}  L  
   =  \frac{ L  \Delta }{ \sqrt{\pi}  \,  \hbar  }  \:
   e^{ - \Delta^2  ( x_\ell )^2  /  \hbar^2  }  \, .
\end{align}

Equation~\eqref{eq:Ex_Prob_App} determines
the condition under which
the uncertainty bound is nontrivial.
The term 
\begin{align}
   \min_{ x_1, x_2,  w_2 }  \Set{
- \log  \left(  \pVH{x_1}  \pVH{x_2}    \,
\Tr \left( \ProjW{w_2}  \right)  \right)  }
\end{align}
dominates the bound~\eqref{eq:Full_Bound}.
The trace equals $2^{ \Sites - 1 }$.
The bound is positive when
$\pVH{x_1}  \pVH{x_2}  2^{ \Sites - 1 }  \leq  1$.
The min, acting on Eq.~\eqref{eq:Ex_Prob_App}, 
chooses $x_1 = x_2 = 0$.
We substitute in from Eq.~\eqref{eq:Ex_Prob_App},
then solve for $L \Delta$:
\begin{align}
   \label{eq:Spin_Chain_Nontrivial_Condn}
   L  \Delta   \leq   \hbar
   \sqrt{  \frac{ \pi }{ 2^{ \Sites - 1 } } }  \, .
\end{align}

Inequality~\eqref{eq:Spin_Chain_Nontrivial_Condn} does not violate
Heisenberg's measurement-disturbance uncertainty relation~\cite{Heisenberg_27_Uber}:
A finite time separates the $\ket{D}$ preparation
from the $\hat{x}$ measurement.
Yet the $\hbar$ and $\frac{1}{ \sqrt{ 2^{ \Sites - 1 } } }$ suggest
that meeting the condition might pose practical difficulties.
A rough estimate offers hope: 
Recent many-body experiments featured
rubidium atoms cooled to 
$\approx 1 \times 10^{ - 5 }$ K~\cite{Bernien_17_Probing}.
The rubidium atom has a mass of
$m  \approx  1  \times  10^{ -25 }$ kg.
Denoting Boltzmann's constant by $\kB$, we approximate
$\kB T  \approx  \frac{ p^2 }{ 2 m }$.
The momentum
$p  \approx  \sqrt{ 2 m \kB T }
\approx  4  \times  10^{ -27 }  \text{ kg $\cdot$ m / s}$
stands in for $\Delta$.
Lengths in periodic arrays can be measured
with X-ray diffraction.
Precisions of up to $L  \approx  10^{ -18 }$ m 
have been achieved with silicon~\cite{Mohr_12_CODATA,Massa_11_Measurement}.
(Though silicon lattices differ from rubidium arrays, 
both numbers reflect precision achievable 
with quantum experiments today.)
Substituting into the bound, then rearranging, yields
$\Sites  \approx  50$.
Approximately the same number of rubidium atoms
formed the quantum simulator in~\cite{Bernien_17_Probing}.


%
%
%
\textbf{Weak-measurement Kraus operators 
$\KrausVH{v_\ell}{j_\ell}  
\equiv  \KrausVH{v_\ell}{x_\ell}$
and coupling strengths $\gVH{j_\ell}  \equiv  \gVH{x_\ell}$:}
The Kraus operators have the form (to within a global phase)
\begin{align}
   \bra{ x_\ell }  \hat{V}_\inter  \ket{ D }  
   =  \braket{ x_\ell }{ D }
   \exp  \left(  -  \frac{ i }{ \hbar }  \,  \tilde{g}  
                        \left[  x_\ell  -  x_0  \right]  \,
                        \ProjVH{ v_\ell }  \right)  \, .
\end{align}
We redefine the Kraus operators such that the coefficient is real:
\begin{align}
   \KrausVH{ v_\ell }{ x_\ell }
   & :=  | \braket{ x_\ell }{ D } |
   \exp  \left(  -  \frac{ i }{ \hbar }  \,  \tilde{g}  
                        \left[  x_\ell  -  x_0  \right]  \,
                        \ProjVH{ v_\ell }  \right)  \\
   \label{eq:Ex_Kraus_Help5_App}
   & =   \sqrt{ \pVH{x_\ell}  } \:  \hat{\id}  
   +  \gVH{x_\ell}   \ProjVH{v_\ell} \, .
\end{align}
The outcome-dependent coupling is
\begin{align}
   \label{eq:Ex_gV_App}  
   \gVH{x_\ell}
   =  \sqrt{ \pVH{x_\ell}  } 
       \left(  e^{ - \frac{i}{ \hbar }  \,  \tilde{g}  
                (  x_\ell  -  x_0  ) }
                - 1  \right)   \, .
\end{align}
We chose $\Delta = 0.1$, which 
(with $L = 0.1$, $\hbar = 1$, and $\Sites = 8$)
satisfies Ineq.~\eqref{eq:Spin_Chain_Nontrivial_Condn}.

\section{Choice of $v_1 = - v_2$ in the spin-chain example}
\label{sec:Why_v1v2} 


Equation~(60) on p. 15 of~\cite{NYH_18_Quasi} motivates our choice.
$\SumKD{\rho} (v_1, w_1, v_2, w_2)$ appears, there, as 
a combination of correlators of $V$ and $\W(t)$.
Let us set $w_1 = w_2$ and replace $\rho$ with $\id$.
We recall that $w_\ell, v_m  =  \pm 1$,
that the Pauli operators' traces vanish,
and that the Pauli operators square to $\id$.
The expression simplifies:
\begin{align}
   \label{eq:Why_v1v2_1}
   \SumKD{\id} (v_1, w_2, v_2, w_2)
   & =  \frac{1}{16}  
   [ 2  +  v_1 v_2  +  2 w_2 ( v_1  +  v_2 )  \expval{ V \W(t) }
     +  ( w_2 )^2  v_1 v_2  \,  F(t)  ]  \, .
\end{align}

Let us analyze the expression piecemeal.
First, the $\expval{ V \W(t) }  \approx  0$ at early times,
because the influence from $V$ has not reached $\W(t)$.
Random-matrix-theory cancellations suppress 
$\expval{ V \W(t) }$ at late times.
Second, the OTOC begins at $F( t \approx 0 )  \approx 1$
and drops to $F( t  \geq  t_* )  \approx  0$.
Third, suppose that $v_1 = - v_2$.

Combining these three behaviors,
we infer the behavior of the RHS of Eq.~\eqref{eq:Why_v1v2_1}.
The first three terms sum to $\approx 1$.
The final term rises from $\approx -1$ to $\approx 0$.
Therefore, $\SumKD{\id} ( v_1, w_2, -v_1, w_2 )$ rises
from $\approx 0$ to $\approx \frac{1}{16}$.

This rise strengthens the bound $f( v_1, v_2{=}{-v_1} )$:
$\SumKD{\id} ( v_1, w_2, -v_1, w_2 )$ contributes to the bound
through the term
\begin{align}
   \label{eq:Why_v1v2_2}
   \frac{ - 2 }{ \ln 2  \:  \Tr \left( \ProjW{w_2} \right)
   \sqrt{  \pV{j_1}  \pV{j_2}  }  }  \:
   \Real  \left(  \gV{j_1} \gV{j_2}  \,
   \SumKD{\id} ( v_1, w_2, -v_1, w_2 )  \right)  
\end{align}
in line~\eqref{eq:Full_Bound_line4}.
We have chosen for the couplings to have 
large imaginary parts, so 
$\gV{j_1} \gV{j_2}$ is dominated by  $- \tilde{g}^2  <  0$.
The quasiprobability is real for all arguments~\cite[p.~24]{NYH_18_Quasi}.
Around $t = t_*$, therefore,~\eqref{eq:Why_v1v2_2} 
rises from $\approx  0$ to 
$\approx \frac{ \tilde{g}^2 }{ 
\ln (2)  \:  2^{ \Sites + 2 }    
\sqrt{  \pV{j_1}  \pV{j_2}  }  }$,
tightening the bound.

In summary, the final $\SumKD{\id}$ value in~\eqref{eq:Full_Bound_line4}
points to $v_1 = -v_2$ as 
a condition under which the uncertainty bound
is relatively tight.
We arbitrarily chose $v_1 = 1$.

Why should the first two $\SumKD{\id}$ terms 
in~\eqref{eq:Full_Bound_line4}
not guide our choice of $(v_1, v_2)$? 
These terms influence the bound through
\begin{align}
   \label{eq:Why_v1v2_3}
   \frac{ -1 }{ \ln 2  \:  \Tr \left( \ProjW{w_2} \right)  }  
   \left[  \frac{ \Verts{  \gV{j_1}  }^2  }{  \pV{j_1 }  }  \:
            \SumKD{\id}  (v_1, w_1, v_1, w_2)
           +  \frac{ \Verts{  \gV{j_2}  }^2  }{  \pV{j_2 }  }  \:
           \SumKD{\id}  (v_2, w_1, v_2, w_2)  \right]  \, .
\end{align}
By Eq.~\eqref{eq:Why_v1v2_1},
$\SumKD{\id} ( v_1, w_1, v_1, w_2 )
=  \SumKD{\id} ( v_2, w_1, v_2, w_2 )
=  \frac{1}{16}  \left[ 3  +  4 w_2  \expval{ V \W(t) }  +  F(t)  \right]$.
As argued earlier, $\expval{ V \W(t) }$ is small
at early and late times.
Hence $\SumKD{\id} (v, w_1, v, w_2)  \geq  0$
for all $v = \pm 1$.
Hence~\eqref{eq:Why_v1v2_3} is expected to be negative,
loosening the bound,
regardless of our choices of $v_1$ and $v_2$.

\section{Calculations: Qubit example for the weak-value uncertainty relation}
\label{app:Qubit_Aw}

$A$ can be weakly measured as follows.
The detector is prepared in the state $\ket{ x+ }$.
A $z$-controlled $y$ conditions a rotation 
of the detector's state on the system's state.
The interaction Hamiltonian 
$H_\inter  =  \tilde{g}  
\left(  \sigma^y_\detect  \otimes  \sigma^z_\sys  \right)$
generates the unitary
\begin{align}
   \label{eq:WVal_Ex_VInter}
   V_\inter  
   & =  \exp \left( - i \tilde{g}  
   \left[  \sigma^y_\detect  \otimes  \sigma^z_\sys  \right]  \right) \\
   & =  \cos ( \tilde{g}  )  \id  
   -  i  \sin  ( \tilde{g} )  
   \left(  \sigma^y_\detect  \otimes  \sigma^z_\sys  \right)  \, .
\end{align}
The detector's $\sigma^y_\detect$ is measured strongly, 
yielding the outcome
$j  =  y_\detect  =  \pm 1$.

%
%
We can begin assembling the ingredients in
Ineq.~\eqref{eq:Weak_Unc_Rel_Rastegin}.
The coupling-free probabilities $\pGen{A}{j}  =  \pGen{Y}{ y_\detect }
=  | \braket{ y_\detect }{ x+ }  |^2  =  \frac{1}{2}$
for $y_\detect  =  \pm 1$.
%
%
Next, we calculate the weak-measurement Kraus operators
$\KrausGenTwo{Y}{j}$
and the outcome-dependent couplings $\gGen{Y}{j}$.

En route to $\KrausGenTwo{Y}{j}  \equiv  \KrausGenTwo{Y}{ y_\detect }$,
we define the physically equivalent
\begin{align}
   \KrausGenTTwo{A}{ y_\detect } 
   =  \bra{ y_\detect }  V_\inter  \ket{ x+ }
   =  \cos ( \tilde{g} )  \braket{ y_\detect }{ x+ }  \id
   - i \sin ( \tilde{g} )  \bra{ y_\detect }  \sigma^y_\detect  \ket{ x+ }  
   \sigma^z_\sys  \, .
\end{align}
We remove a global phase:
\begin{align}
   \KrausGenTwo{A}{j}  
   & =  \frac{ | \braket{ y_\detect }{ x+ } | }{ \braket{ y_\detect }{ x+ } }  \:
   \KrausGenTTwo{Y}{ y_\detect }  \\
   & =  \cos ( \tilde{g} )  \:  |  \braket{ y_\detect }{ x+ } |  \:  \id
   - i \sin ( \tilde{g} )  \,
   \frac{  \bra{ y_\detect } \sigma^y_\detect  \ket{ x+ }  }{  
             \braket{ y_\detect }{ x+ }  }  \,
   | \braket{ y_\detect }{ x+ } |  \,
   \sigma^z_\sys  \, .
\end{align}
To first order in $\tilde{g}$,
\begin{align}
   \label{eq:WVal_Ex_KrausB}
   \KrausGenTwo{Y}{ y_\detect }  
   =  \sqrt{  \pGen{Y}{ y_\detect }  }  \,  \id
   +  \gGen{Y}{ y_\detect }  \:  \sigma^z_\sys
   +  O  \left(  \tilde{g}^2  \right)  \, .
\end{align}
The outcome-dependent coupling has the form
\begin{align}
   \gGen{A}{j}
   \equiv   \gGen{Y}{ y_\detect } 
   :=  - i \tilde{g}  \bra{ y_\detect }  \sigma^y  \ket{ x+ }  \:
   \frac{ | \braket{ y_\detect }{ x+ }  | }{  \braket{ y_\detect }{ x+ } }  \,
   =  \frac{ - y_\detect  \,  i  \: }{ \sqrt{2} }  \,  \tilde{g}  \, .
\end{align}

The weak value has the form
$A_\weak ( z_\sys ,  x_\sys )  
   =  \bra{ x_\sys }  \sigma^y_\sys  \ket{ z_\sys }  /
   \braket{ x_\sys }{ z_\sys }
   =  x_\sys  \,  z_\sys  \,  i$.
The nonreality is nonclassical~\cite{Dressel_12_Significance}.

Let us calculate the bound~\eqref{eq:Bound_Beyond}.
$f_\weak$ contains a factor
$\Real  \left(  \gGen{A}{j}  \:  A_\weak (i, f)  \right)
=  \Real  \left(  \gGen{Y}{ y_\detect }  \:  
                    A_\weak ( z_\sys ,  x_\sys )  \right)
=  y_\detect  \,  x_\sys  \,  z_\sys  \,  \tilde{g}  /  \sqrt{2}$.
When this factor maximizes at
$\frac{ | \tilde{g} | }{ \sqrt{2} }$,
the minimum in $f_\weak$ is attained.
The probability
$\Tr  \left(  \Pi^F_f  \Pi^\I_i  \right)
=  | \braket{ x_\sys }{ z_\sys } |^2  =  \frac{1}{2}$
for all $x_\sys, z_\sys  = \pm 1$.
Substituting into Eq.~\eqref{eq:Bound_Beyond} yields
\begin{align}
   & \min_{ z_\sys, y_\detect, x_\sys }  \Bigg\{
   - \log  \left(  \pGen{Y}{ y_\detect }  |  \braket{ x_\sys }{ z_\sys }  |^2  \right)
   %
   -  \frac{2}{ \ln 2  \:  \sqrt{  \pGen{Y}{ y_\detect }  }  }  \:
   \Real  \left(  \gGen{Y}{ y_\detect }  \:
   A_\weak ( z_\sys, x_\sys )  \right)
   + O \left(  \tilde{g}^2  \right)  \Bigg\}
   =  2  -  \frac{2}{ \ln 2 }  \,  | \tilde{g} |
   + O  \left(  \tilde{g}^2  \right)   \, .
\end{align}

%
%
Having evaluated the RHS of Ineq.~\eqref{eq:Weak_Unc_Rel_Rastegin},
we turn to the LHS.
We calculate the POVM probabilities,
then their entropies.
$\rho$ denotes an arbitrary system state,
exemplified by $\ket{ z + }$.

POVM II consists of a strong $\I  =  \sigma^z_\sys$ measurement.
The possible outcomes $z_\sys$ have probabilities
$q^{\rm II}_{z_\sys} 
   =  \bra{ z_\sys }  \rho  \ket{ z_\sys }$
of obtaining.
If $\rho = \ketbra{ z+ }{ z+ }$, then
$q^{\rm II}_{z_\sys = 1}  =  1$, and
$q^{\rm II}_{z_\sys = -1}  =  0$.
The max entropy is $H_\Max \left(  \Set{  q^{\rm II}_{z_\sys}  }  \right)
=  \log 1  =  0$.
Smoothing cannot change this value.

POVM I consists of 
a weak $A = \sigma^y_\sys$ measurement
followed by a strong $F = \sigma^x_\sys$ measurement.
The possible outcome tuples $(y_\detect, x_\sys)$
correspond to the probabilities
$q^{\I}_{ y_\detect, x_\sys } 
   =  \bra{ x_\sys }  \KrausGenBB{Y}{ y_\detect } \rho 
   \left(  \KrausGenBB{Y}{ y_\detect }  \right)^\dag 
   \ket{ x_\sys }  \, .$
We substitute in, 
then multiply out:
\begin{align}
   q^{\I}_{ y_\detect, x_\sys } 
   & =  \pGen{Y}{ y_\detect }  \bra{ x_\sys }  \rho  \ket{ x_\sys }
   +   \sqrt{ \pGen{Y}{ y_\detect }  }
   \Big[  \gGen{Y}{ y_\detect }   \bra{ x_\sys }  \sigma^z  \rho  \ket{ x_\sys }
             +  \left(  \gGen{Y}{ y_\detect }  \right)^*
                 \bra{ x_\sys }  \rho  \sigma^z  \ket{ x_\sys }  \Big]
   +  O  \left(  \tilde{g}^2  \right)  \, .
\end{align}
If $\rho  =  \ketbra{ z+ }{ z+ }$,
the distribution is uniform: 
$q^{\I}_{ y_\detect, x_\sys }
=  \frac{1}{4}  +  O  \left(  \tilde{g}^2  \right)$
for all $y_\detect,  x_\sys  =  \pm 1$.
Hence $H_\Min \left(  \Set{  q^{\I}_{ y_\detect, x_\sys }  }  \right)
=  2.00  +  O  \left(  \tilde{g}^2  \right)$.
Nor can smoothing alter this value.

\end{appendices}

%
%
\bibliographystyle{h-physrev}
\bibliography{OTOC_bib}


\end{document}